\documentclass[a4paper,11pt]{article}
\usepackage{jheppub}

\usepackage[utf8]{inputenc}
\usepackage{amsmath}
\usepackage{amssymb}
\usepackage{mathrsfs}
\usepackage{graphicx}
\usepackage{bbold}
\usepackage{comment}
\usepackage[dvipsnames]{xcolor}
\usepackage{braket}
\usepackage{bm,bbm}
\usepackage{physics}
\usepackage{tikz}

\newcommand\beq{ \begin{eqnarray} }
\newcommand\eeq{ \end{eqnarray} }

\usepackage{bbm}

\preprint{YITP-24-68, RIKEN-iTHEMS-Report-24}
\arxivnumber{1234.56789} % if you have one

\title{ Lattice study on finite density QC$_2$D towards zero temperature}

\author[a]{Kei Iida,}
\author[b,c]{Etsuko Itou,}
\author[c,d]{Kotaro~Murakami,}
\author[e,f]{and Daiki~Suenaga}
\affiliation[a]{Department of Mathematics and Physics, Kochi University, 2-5-1 Akebono-cho, Kochi 780-8520, Japan }%\email{iida@kochi-u.ac.jp}}  

\affiliation[b]{Yukawa Institute for Theoretical Physics, Kyoto University, Kitashirakawa Oiwakecho, Sakyo-ku, Kyoto 606-8502, Japan }
\affiliation[c]{Interdisciplinary Theoretical and Mathematical Sciences Program (iTHEMS), RIKEN, Wako, Saitama 351-0198, Japan}

\affiliation[d]{Department of Physics, Tokyo Institute of Technology, 2-12-1 Ookayama, Megro, Tokyo 152-8551, Japan }
\affiliation[e]{Kobayashi-Maskawa Institute for the Origin of Particles and the Universe, Nagoya University, Nagoya, 464-8602, Japan}
\affiliation[f]{Research Center for Nuclear Physics,
Osaka University, Ibaraki 567-0048, Japan }

\emailAdd{iida@kochi-u.ac.jp}
\emailAdd{itou@yukawa.kyoto-u.ac.jp}
\emailAdd{kotaro.murakami@yukawa.kyoto-u.ac.jp}
\emailAdd{suenaga.daiki.j1@f.mail.nagoya-u.ac.jp}

\abstract{
We investigate the phase structure and the equation of state (EoS) for dense two-color QCD (QC$_2$D) at low temperature ($T = 40$ MeV, $32^4$ lattice) for the purpose of extending our previous works~\cite{Iida:2019rah, Iida:2022hyy} at $T=80$ MeV ($16^4$ lattice).  Indeed, a rich phase structure below the pseudo-critical temperature $T_c$ as a function of quark
chemical potential $\mu$  has been revealed, but finite volume effects in a high-density regime sometimes cause a wrong understanding. Therefore, it is important to investigate the temperature dependence down to zero temperature with large-volume simulations. 
By performing $32^4$ simulations, we obtain essentially similar results to the previous ones, but we are now allowed to get a fine understanding of the phase structure via the temperature dependence.
Most importantly, we find that the hadronic-matter phase, which is composed of thermally excited hadrons, shrinks with decreasing temperature and that the diquark condensate scales as $\langle qq \rangle \propto \mu^2$ in the BCS phase, a property missing at $T=80$ MeV.
From careful analyses, furthermore, we confirm a tentative conclusion that the topological susceptibility is independent of $\mu$.
We also show the temperature dependence of the pressure, internal energy, and sound velocity as a function of $\mu$. The pressure increases around the hadronic-superfluid phase transition more rapidly at the lower temperature, while the temperature dependence of the sound velocity is invisible. Breaking of the conformal bound is also confirmed thanks to the smaller statistical error.
}

\begin{document}
\maketitle
\flushbottom
%%%%%%%%%%%%%%%%%%%%%%%%%%%%%%%%%%%%%%%%%%%%%%%%%%%%%%
%%%%%%%%%%%%%%%%%%%%%%%%%%%%%%%%%%%%%%%%%%%%%%%%%%%%%%
\section{Introduction}
\label{sec:intro}
%%%%%%%%%%%%%%%%%%%%%%%%%%%%%%%%%%%%%%%%%%%%%%%%%%%%%%
%%%%%%%%%%%%%%%%%%%%%%%%%%%%%%%%%%%%%%%%%%%%%%%%%%%%%%
Equilibrium phase structure of QCD in a broad range of baryon densities and temperatures underlies the physics of the early universe, stellar collapse, neutron stars, and highly energetic heavy-ion collisions.  The regime of high densities and low temperatures is of particular importance because the equation of state (EoS) of matter in neutron stars is well constrained from recent neutron star observations via X-rays and gravitational waves.  The significance of the phase structure increases more and more as new data are expected from neutron star observations and intermediate-energy heavy-ion collision experiments in the near future.  Theoretically, however, the phase diagram is still elusive except for the limiting cases of nearly zero and asymptotically large baryon chemical potential.

The action of finite-density QCD in the continuum Euclidean spacetime is given by
\beq
S = \int d^4x \left[\frac{1}{2} \mathrm{tr} F_{\mu \nu} F_{\mu \nu} + \bar{\psi}(x) (\gamma_\nu D_\nu + m +\mu \gamma_4) \psi(x)\right].
\eeq
Here, $\mu$ denotes the quark chemical potential. Introducing this quark chemical potential term causes the infamous sign problem in the first-principles lattice Monte Carlo simulation.
The sign problem has been proven to be an NP-hard problem in general~\cite{Troyer:2004ge}, so that an alternative approach is necessary to investigate the high-density QCD(-like) theory (see the recent review paper~\cite{Nagata:2021ugx} for the sign problem in dense QCD).
Furthermore, at sufficiently low temperature and high chemical potential, particularly $\mu \geq m_{\pi}/2$  with the pion mass 
$m_\pi$, another numerical problem emerges.   The chemical potential acts to spread an eigenvalue distribution of the Dirac operator in the complex plane. It is known that such an eigenvalue distribution makes it extremely hard to perform the Hybrid Monte Carlo simulation. It is related to the so-called (early-)onset problem and comes from the generation of the lightest hadron (i.e., pion) pairs under $\mu \geq m_{\pi}/2$~\cite{Barbour:1991vs, Barbour:1997bh, Barbour:1997ej, Barbour:1999mc, Cohen:2003kd, Adams:2004yy, Nagata:2012ad}.

One promising approach is the first-principles study of two-color QCD (QC$_2$D) for qualitative understanding of the high density regime of QCD-like theories~\cite{Muroya:2000qp, Muroya:2002jj, Muroya:2002ry}. It is well-known that QC$_2$D at $\mu=0$ shares the nonperturbative properties with real QCD, such as confinement, chiral symmetry breaking, and emergence of instantons.
 In contrast to real QCD, on the other hand, QC$_2$D is free from the sign problem if we consider an even-flavor theory; the fundamental representation of the SU($2$) group, to which quark fields belong, is pseudo-real.
Furthermore, if we introduce the diquark source term~\cite{Kogut:2001na},
\beq
S_{\mathrm{diquark}}= \int d^4x \left[-j \bar{\psi}_1 (C \gamma_5) \tau_2 \bar{\psi}_2^{T} + \bar{j} \psi_2^T (C \gamma_5) \tau_2 \psi_1\right], \label{eq:diquark-source-cont}
\eeq
then, it is possible to perform the simulation beyond the onset problem and hence obtain the overall QC$_2$D phase diagram on $T$--$\mu$ plane.
Here, we consider $N_f=2$ case. The indices $1, 2$ denote the flavor label.
The additional parameter, $j$ and $\bar{j}$,  correspond to the anti-diquark and diquark source parameters, respectively.
This term breaks the diquark-antidiquark symmetry explicitly. By taking the $j \rightarrow 0$ limit, we can address the spontaneous breaking of this symmetry.

In recent years, lattice Monte Carlo studies on dense QC$_2$D have been conducted independently and intensively by several groups~\cite{Hands:2006ve, Hands:2007uc, Hands:2011hd, Cotter:2012mb, Cotter:2012ny, Hands:2012fs, Boz:2013rca, Boz:2015ppa,Braguta:2016cpw, Itou:2018vxf, Astrakhantsev:2018uzd, Boz:2019enj, Iida:2019rah, Astrakhantsev:2020tdl, Buividovich:2020dks, Iida:2020emi, Ishiguro:2021yxr, Begun:2022bxj, Murakami:2022lmq, Murakami:2023ejc, Iida:2022hyy, Itou:2022ebw, Lawlor:2022hfi, Itou:2023pcl}.
Putting the results from Refs.~\cite{Cotter:2012mb, Cotter:2012ny, Braguta:2016cpw, Iida:2019rah, Boz:2019enj, Begun:2022bxj} together, one can conclude that the QC$_2$D phase diagram is not only qualitatively  but also quantitatively clarified.
Most remarkably, it has been found that diquark condensation, which is responsible for the emergence of superfluidity, remains at temperature of order or even higher than $100$ MeV.
In such a superfluid phase, several lattice simulations exhibit a Bose-Einstein condensation (BEC) to BCS crossover as the baryon density increases.
Furthermore, independent lattice studies~\cite{Boz:2019enj, Begun:2022bxj, Ishiguro:2021yxr} show that confinement remains at temperatures below $100$ MeV even in the high-density region ($\mu > 1$ GeV), while some local quantities, such as the quark number density and diquark condensate, can be understood in terms of Fermi-degenerate deconfined quarks.
Thus, it is natural to consider that
a kind of hadronic superfluidity occurs in low-temperature and high-density QC$_2$D.

Another important result is the new finding that the squared speed of sound exceeds the conformal bound, namely $c_{\rm s}^2/c^2 >1/3$, around the crossover regime of the superfluid phase~\cite{Iida:2022hyy}.
The EoS and sound velocity in dense QCD at low temperature have attracted much attention recently;
discussions on whether the speed of sound actually exceeds the conformal bound have been intensively done via effective model analyses and observational data for neutron star masses and radii~\cite{Masuda:2012ed,Baym:2017whm, McLerran:2018hbz,Fujimoto:2020tjc, Kojo:2021ugu, Kojo:2021hqh, Braun:2022jme, Brandes:2023hma, Kawaguchi:2024iaw}.
The result from our previous work~\cite{Iida:2022hyy} presents the first evidence for $c_{\rm s}^2/c^2 >1/3$ among existing Monte Carlo simulations of QCD-like theories.
After our work, two independent lattice calculations for three-color QCD with isospin chemical potential have succeeded in finding such an excess of the speed of sound~\cite{Brandt:2022fij, Brandt:2022hwy, Abbott:2023coj}.

Bearing these situations in mind, in this work, we investigate the phase structure and the EoS for dense QC$_2$D at even lower temperature. We have studied such equilibrium properties at $T = 80$ MeV ($16^4$ lattice) in our previous papers~\cite{Iida:2019rah, Iida:2022hyy}, where the emergence of the superfluid phase and the breaking of the conformal bound have been already observed. Here, we aim to newly add lower temperature data by using $32^4$ lattices, corresponding to $T = 40$ MeV, and clarify the temperature dependence of the physical quantities in the superfluid phase.
At $\mu=0$, there are only quark-gluon-plasma (QGP) and hadronic phases depending on the temperature. Then, the simulation on the hypercubic lattice ($N_s = N_\tau$) with sufficiently large $N_s$, which corresponds to a temperature below $T_c$, is called the ``zero-temperature” simulation conventionally.
At non-zero $\mu$, however, it has been revealed that various phases emerge below $T_c$ in the finite density region and that finite volume effects and some lattice artifacts are sometimes large~\cite{Boz:2019enj, Begun:2022bxj, Lawlor:2022hfi}. Consequently, it is important to
investigate how the physical quantities depend on the temperature as the temperature goes to zero by performing large volume simulations.
As for the phase diagram, we will particularly investigate the fate of the hadronic-matter phase  composed of thermally excited hadrons, 
 the scaling behavior of the diquark condensate, and the $\mu$-(in)dependence of the topological susceptibility.
For the EoS and sound velocity, we will address the negativity of the trace anomaly and the breaking of the conformal bound by comparing
with recent related works.

This paper is organized as follows. 
In Section~\ref{sec:setup}, we explain our lattice action and simulation parameters. 
We present the simulation results for the phase diagram at $T=40$ MeV in Section~\ref{sec:phase} and for the EoS and sound velocity in Section~\ref{sec:eos}.  For comparison,
we give a brief summary of recent related works relevant to the sound velocity of QCD(-like) theories in Sec.~\ref{sec:conformal-bound}. 
Section~\ref{sec:summary} is devoted to the summary. Furthermore, we tabulate the raw data for the EoS and sound velocity in Appendix~\ref{sec:raw-data}.

%%%%%%%%%%%%%%%%%%%%%%%%%%%%%%%%%%%%%%%%%%%%%%%%%%%%%%
%%%%%%%%%%%%%%%%%%%%%%%%%%%%%%%%%%%%%%%%%%%%%%%%%%%%%%
\section{Lattice setup}\label{sec:setup}
%%%%%%%%%%%%%%%%%%%%%%%%%%%%%%%%%%%%%%%%%%%%%%%%%%%%%%
%%%%%%%%%%%%%%%%%%%%%%%%%%%%%%%%%%%%%%%%%%%%%%%%%%%%%%

%%%%%%%%%%%%%%%%%%%%%%%%%%%%%%%%%%%%%%%%%%%%%%%%%%%%%%
%%%%%%%%%%%%%%%%%%%%%%%%%%%%%%%%%%%%%%%%%%%%%%%%%%%%%%
\subsection{Lattice action}
\label{subsec:action}
%%%%%%%%%%%%%%%%%%%%%%%%%%%%%%%%%%%%%%%%%%%%%%%%%%%%%%
%%%%%%%%%%%%%%%%%%%%%%%%%%%%%%%%%%%%%%%%%%%%%%%%%%%%%%
As a lattice gauge action relevant to QC$_2$D, we utilize the Iwasaki gauge action~\cite{Iwasaki:1983iya}, which is composed of the plaquette term with $W^{1\times 1}_{\mu\nu}$ 
and the rectangular term with $W^{1\times 2}_{\mu\nu}$,  
\beq
S_g = \beta \sum_x \left(
 c_0 \sum^{4}_{\substack{\mu<\nu \\ \mu,\nu=1}} W^{1\times 1}_{\mu\nu}(x) +
 c_1 \sum^{4}_{\substack{\mu\neq\nu \\ \mu,\nu=1}} W^{1\times 2}_{\mu\nu}(x) \right) 
\eeq
with $\beta=4/g_0^2$, where $g_0$ denotes the bare 
gauge coupling constant.  The coefficients $c_0$ and $c_1$ are set to 
 $c_0=1-8c_1$ with $c_1=-0.331$.

As a lattice fermion action applicable to dense QC$_2$D, we use the two-flavor Wilson fermion action, 
\beq
S_F= \bar{\psi}_1 \Delta(\mu)\psi_1 + \bar{\psi}_2 \Delta(\mu) \psi_2.\label{eq:fermion-action-wo-j}
\eeq
Here, $\Delta(\mu)$ denotes the Wilson-Dirac operator including the number operator, 
\begin{align}
    \Delta(\mu)_{x,y} = \delta_{x,y} 
&- \kappa \sum_{i=1}^3  \left[ ( \mathbb{I}_4 - \gamma_i)  U_{x,i}\delta_{x+\hat{i},y} + (\mathbb{I}_4+\gamma_i)  U^\dagger_{y,i}\delta_{x-\hat{i},y}  \right] \nonumber\\
&- \kappa   \left[ e^{+\mu}( \mathbb{I}_4 - \gamma_4)  U_{x,4}\delta_{x+\hat{4},y} + e^{-\mu}(\mathbb{I}_4+\gamma_4)  U^\dagger_{y,4}\delta_{x-\hat{4},y}  \right],\label{eq:Dirac-op}
\end{align}
where the chemical potential is incorporated via an exponential factor, rather than a linear coefficient of $\gamma_4$, to avoid a quadratic divergence in the continuum limit~\cite{Hasenfratz:1983ba}, and $\kappa$ is the hopping parameter.
This $\Delta(\mu)$ breaks the $\gamma_5$-hermiticity, but still satisfies $\Delta(\mu)^\dag = \gamma_5 \Delta(-\mu) \gamma_5$. In the lattice Monte Carlo simulation, the fermion action is expressed as $\det (\Delta (\mu))$ by integrating out the fermion fields beforehand.
At $\mu \ne 0$, 
\beq
(\det \Delta (\mu))^* = \det \Delta (\mu) ^\dag = \det \Delta (-\mu) \label{eq:non-hermicity}
\eeq
implies that the fermion action takes  a complex value in general.

In the case of SU($2$) gauge, however, the link variable is $U_\mu= e^{i A_\mu^a \tau^a}$, which satisfies
\beq
U_{\mu}^* = \tau_2 U_\mu \tau_2,
\eeq
where $\tau_2$ denotes the Pauli matrix and acts on the color indices.
Then, for given $\mu$, the Dirac operator has the following conjugate property:
\beq
\Delta (U,\gamma_\mu)^* = \Delta (U^*, \gamma_\mu^*)= \tau_2 C \gamma_5 \Delta (U,\gamma_\mu) (C \gamma_5)^{-1} \tau_2,
\eeq
where $C \gamma_5$ acts on the spinor indices, and $C$ denotes the charge conjugation operator; here we take $C=i \gamma_0 \gamma_2$.
Consequently, in addition to 
Eq.~\eqref{eq:non-hermicity},  
one obtains
\beq
[\det \Delta (U,\gamma_\mu)]^* = \det [\tau_2 C \gamma_5 \Delta (U,\gamma_\mu) (C \gamma_5)^{-1} \tau_2 ]= \det \Delta (U,\gamma_\mu).
\eeq
Thus, the fermion action in SU($2$) gauge theory takes a real value.
Essentially, this comes from the pseudo-reality of the fundamental representation, namely ``quarks'', of the SU($2$) group.
Note that here the positivity of the fermion action is not guaranteed. If we consider an odd-number fermion system, the action can take a real but negative value, and hence the sign problem appears even in QC$_2$D.

To perform the low-temperature and high-density QC$_2$D simulations, the reality of the fermion action is necessary but insufficient.
In general QCD(-like) theory, it is known that a numerical instability related to the (early-)onset problem appears around $\mu = m_{\pi}/2$ . In the case of QC$_2$D, in such a $\mu$ regime occurs a phase transition from the hadronic (normal) to superfluid phase.
Then, the diquark mass becomes so small that
the simulation could not proceed beyond $\mu \approx m_{\pi}/2$ even if we utilized a very tiny molecular-dynamics step ($\sim 1/1000$) in the Hybrid Monte Carlo algorithm.

Therefore, we introduce the diquark source term, Eq.~\eqref{eq:diquark-source-cont}, which leads to the fermion action on the lattice,
\beq
S_F= \bar{\psi}_1 \Delta(\mu)\psi_1 + \bar{\psi}_2 \Delta(\mu) \psi_2 - J \bar{\psi}_1 (C \gamma_5) \tau_2 \bar{\psi}_2^{T} + \bar{J} \psi_2^T (C \gamma_5) \tau_2 \psi_1,\label{eq:action}
\eeq
with $J=j \kappa$ and $\bar{J}= \bar{j} \kappa$ corresponding to the anti-diquark and diquark source parameters, respectively. Here, the factor $\kappa$ comes from the rescaling of the Wilson fermion on the lattice. 
For simplicity, we put $J=\bar{J}$ and assume that it takes a real value.

The action~\eqref{eq:action} can be rewritten by using an extended fermion matrix ($\mathcal M$) as
\beq
S_F&=& (\bar{\psi}_1 ~~ \bar{\varphi}) \left( 
\begin{array}{cc}
\Delta(\mu) & J \gamma_5 \\
-J \gamma_5 & \Delta(-\mu) 
\end{array}
\right)
\left( 
\begin{array}{c}
\psi_1  \\
\varphi  
\end{array}
\right)
 \equiv  \bar{\Psi} {\mathcal M} \Psi,  \label{eq:def-M}
\eeq
where
$\bar{\varphi}=-\psi_2^T C \tau_2, ~~~ \varphi=C^{-1} \tau_2 \bar{\psi}_2^T.$
Thanks to the pseudo-reality of the fundamental fermions in the SU($2$) gauge theory, $\psi_1$ and the charge conjugation of $\bar{\psi}_2^T$ can be put in the same multiplet. 
The square of the extended matrix can be diagonal if $J(=\bar{J})$ is real.   
We thus obtain
\beq
\det[{\mathcal M}^\dag {\mathcal M}] = 
\det[ \Delta^\dag(\mu)\Delta(\mu) + J^2 ]~\det[  \Delta^\dag (-\mu) \Delta(-\mu) + J^2  ]. \label{eq:MdagM}
\eeq
From the point of view of actual numerical calculations, the $J$ insertion speeds up calculations since it lifts the eigenvalues of the matrix up~\cite{Kogut:2001na, Kogut:2002cm,Skullerud:2003yc}.

In fact, $\det[{\mathcal M}^\dag {\mathcal M}]$ corresponds to the fermion 
action for the four-flavor theory, since a single $\mathcal{M}$ in 
Eq.~\eqref{eq:def-M} represents the fermion kernel of the two-flavor 
theory.  To reduce the number of flavors, we take the square root of 
the extended matrix in the action and utilize the Rational Hybrid Monte Carlo (RHMC) algorithm in our numerical simulations.

%%%%%%%%%%%%%%%%%%%%%%%%%%%%%%%%%%%%%%%%%%%%%%%%%%%%%%
%%%%%%%%%%%%%%%%%%%%%%%%%%%%%%%%%%%%%%%%%%%%%%%%%%%%%%
\subsection{Simulation parameters}
\label{subsec:para}
%%%%%%%%%%%%%%%%%%%%%%%%%%%%%%%%%%%%%%%%%%%%%%%%%%%%%%
%%%%%%%%%%%%%%%%%%%%%%%%%%%%%%%%%%%%%%%%%%%%%%%%%%%%%%
We perform the simulations with ($\beta, \kappa, N_s, N_\tau$) $=$ 
($0.800,0.159,32,32$), which are complementary to our previous work where ($N_s, N_\tau$) $=$ ($16,16$) and ($32,8$) lattices with the same $\beta$ and $\kappa$ were utilized~\cite{Iida:2019rah, Iida:2022hyy}.
Although QC$_2$D is not a real theory, we fix the ``physical'' scale so that the scale correspondence is easy to understand.
According to Ref.~\cite{Iida:2020emi}, the chiral susceptibility at $\mu=0$ 
 is peaked at $(\beta, N_\tau$) $=$ ($0.950,10$) with the simulations performed at fixed $m_{\rm PS}/m_{V}$, where $m_{\rm PS}$  and $m_V$ are the pseudo-scalar and vector meson masses in vacuum ($\mu=0$), respectively~\footnote{Hereafter, we denote the ``pion'' and ``rho meson'' in 2-color QCD as the pseudoscalar (PS) and vector (V) mesons, respectively, to avoid confusion with the ordinary pion and rho meson in 3-color QCD.}. 
Using the scale setting with $w_0$~\cite{BMW:2012hcm} scale in the gradient flow method~\cite{Luscher:2010iy}, we found that ($\beta, N_\tau$) $=$ ($0.800, 32$) simulation corresponds to $T = 0.19 T_c = 40$ MeV with $a=0.17$ fm if we set $T_c = 200$ MeV to fix a physical unit.

In terms of the hadronic quantities, the simulations with ($\beta,\kappa$) $=$ ($0.800, 0.159$) give $m_{\rm PS}/m_V=0.813(1)$ and $am_{\rm PS}= 0.620(1)$ ($m_{\rm PS}\approx 738$ MeV) ~\cite{Murakami:2022lmq, Murakami:2023ejc, Murakami-progress}. 
These values, which are obtained by using $400$ configurations, are updated from our previous work~\cite{ Iida:2020emi}, where we obtained $am_{\rm PS}= 0.623(3)$ with less statistics.
Either of the $m_{\rm PS}$ values will be utilized to make the chemical potential dimensionless like $\mu/m_{\rm PS}$.
In this manuscript, when comparing the present data at $T=40$ MeV ($32^4$ lattice) and the previous data~\cite{Iida:2019rah, Iida:2022hyy} at $T=80$ MeV ($16^4$ lattice), the present and previous data are  plotted as a function of the chemical potential normalized by $am_{\rm PS}=0.620$ and 
$am_{\rm PS}=0.623$, respectively.  As we will see later, 
replacement of 0.623 with 0.620 in plotting the previous data would
lead to just a tiny shift of the horizontal axis.

As for the chemical potential
$\mu$, we take $0 \le a\mu \le 0.75$ at intervals of $\Delta a\mu =0.05$ in this work. We also investigate the case of $a\mu =0.27$ closely to see the phase transition between the hadronic and superfluid phases.
The upper bound of $\mu$, $a\mu \le 0.75$, is determined in such a way that lattice artifacts are suppressed.
In fact, a peculiar phase is induced by the lattice artifacts at sufficiently high density as shown in the previous work~\cite{Iida:2019rah}.
There are two possible origins of the lattice artifacts that are responsible for emergence of such a phase in the numerical 
simulations. 
The first one can be found by observing the behavior of the diquark condensate. In our previous work~\cite{Iida:2019rah}, we observed that the amplitude of $\langle qq \rangle$ decreases with increasing $\mu$
in the range of $a\mu \gtrsim 0.80$ where
a partial suppression of quark propagation occurs in the Dirac operator, Eq.~\eqref{eq:Dirac-op}. A similar behavior was 
reported in Refs.~\cite{Kogut:2002cm, Braguta:2016cpw} in the $N_f=2,4$
simulations using the staggered fermions.
The second one
is related to the finite volume effect. Once the quark number density $n_q$ becomes sufficiently high that all the lattice sites are occupied,
the simulation results must be incorrect. 
The signal of such a situation can be found as a saturation of $n_q$ in lattice unit, which corresponds to $a^3 n_q=2N_cN_f=8$ in our notation.
In the actual simulation of the present work, the maximal value of $ a^3 \langle n_q \rangle$ is  $0.10$ at $(a\mu, aj) = (0.75, 0.02)$, 
which is significantly smaller than $8$.  We can thus safely avoid these artifacts in our simulation.

As for the range of $aj$, we can perform the configuration generation at $aj=0$ for $a\mu \leq 0.25$, while we have to introduce non-zero $aj$ for $a\mu \geq 0.27$ to avoid the numerical instability problem. We take $aj = 0.010, 0.015$, and $0.020$ in the case of $a\mu \geq 0.27$.
After measuring physical observables at each $aj$, we take the $j \rightarrow 0$ limit for each fixed value of $\mu$.
In the present work where we generate the configurations at the above three values of $aj$ for ($N_s, N_\tau$)$=$ (32,32), we can improve the $j \rightarrow 0$ extrapolation over the previous work where 
the configuration generations were performed only for $j=0.010$ and $0.020$ 
for ($N_s, N_\tau$)$=$ (16,16)~\cite{Iida:2019rah, Iida:2022hyy}.

Indeed, we have generated $100$ configurations in the simulations for each parameter.
The statistical errors are estimated by the jackknife methods.
Although the number of the generated configurations for each parameter is almost the same as that in the previous work on $16^4$ lattices,
the statistical errors do decrease. This is because most of the observables of interest here, which are given by local operators on the lattice, have been measured by taking an average over all lattice sites.

%%%%%%%%%%%%%%%%%%%%%%%%%%%%%%%%%%%%%%%%%%%%%%%%%%%%%%
%%%%%%%%%%%%%%%%%%%%%%%%%%%%%%%%%%%%%%%%%%%%%%%%%%%%%%
\section{Phase structure}\label{sec:phase}
%%%%%%%%%%%%%%%%%%%%%%%%%%%%%%%%%%%%%%%%%%%%%%%%%%%%%%
%%%%%%%%%%%%%%%%%%%%%%%%%%%%%%%%%%%%%%%%%%%%%%%%%%%%%%

%%%%%%%%%%%%%%%%%%%%%%%%%%%%%%%%%%%%%%%%%%%%%%%%%%%%%%%%%%%%%%%%%%%%%%%%%
%%%%%%%%%%%%%%%%%%%%%%%%%%%%%%%%%%%%%%%%%%%%%%%%%%%%%%%%%%%%%%%%%%%%%%%%%
\subsection{Observables and definition of phases}
\label{subsec:obs}
%%%%%%%%%%%%%%%%%%%%%%%%%%%%%%%%%%%%%%%%%%%%%%%%%%%%%%%%%%%%%%%%%%%%%%%%%
%%%%%%%%%%%%%%%%%%%%%%%%%%%%%%%%%%%%%%%%%%%%%%%%%%%%%%%%%%%%%%%%%%%%%%%%%

\begin{figure}[htbp]
\centering
\includegraphics[width=0.8 \textwidth]{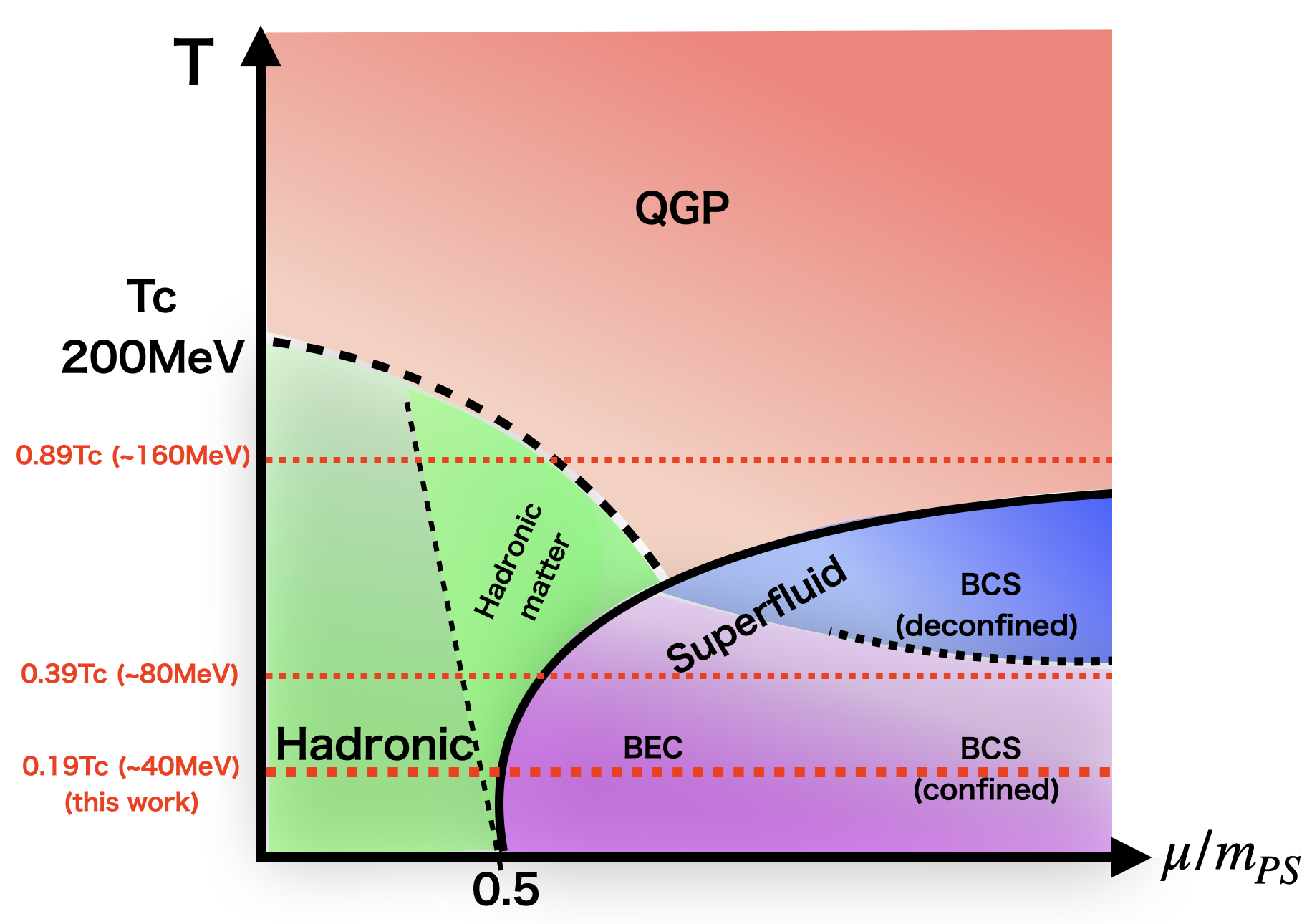}
\caption{Schematic QC$_2$D phase diagram. In our previous work~\cite{Iida:2019rah}, we clarified the phase structure at $T=160$ MeV and $80$ MeV, while in this work, we newly address what it is like at $T=40$ MeV.}\label{fig:schematic}
\end{figure}

According to our previous work~\cite{Iida:2019rah} and related papers~\cite{Hands:2006ve, Braguta:2016cpw, Boz:2019enj}, one can summarize the phase diagram on $T$--$\mu$ plane as shown in Figure~\ref{fig:schematic}.  
%%%%%%%%%%%%%%%%%%%%%%%%%%%%%%%%%%%%%%%%%%%%%%%%%%%%%%%%%%
%%%%%%%%%%%%%%%%%%%%%%%%%%%%%%%%%%%%%%%%%%%%%%%%%%%%%%%%%%
We use the name of each phase as shown in Table~\ref{table:phase}.
Here, we utilize three quantities, the magnitude of the Polyakov loop ($\langle |L| \rangle$), the diquark condensate ($\langle qq \rangle$), and the quark number density ($\langle n_q \rangle$) to distinguish among different phases.
%%%%%%%%%%%%%%
\begin{table}[h]
\begin{center}
\begin{tabular}{|c||c|c|c|c|c|}
\hline
 \multicolumn{1}{|c||}{}  & \multicolumn{2}{c|}{Hadronic } &     \multicolumn{2}{c|}{Superfluid } &\multicolumn{1}{c|}{QGP} \\  
\cline{3-3} \cline{4-5}  & & Hadronic matter ($T>0$) &BEC & BCS  &  \\  
 \hline \hline
$\langle |L| \rangle$ & zero  & zero  &   &   & non-zero \\
$\langle qq \rangle$ & zero  &  zero  & non-zero & ($\propto \mu^2$)  & zero \\ 
$ \langle n_q \rangle $ &   zero &  non-zero  & non-zero & $\langle n_q \rangle /n_q^{\rm tree} \approx 1$ & non-zero \\ 
 \hline
\end{tabular}
\caption{ Definition of the phases. To distinguish between the BEC and BCS phases, we use the value of $\langle n_q \rangle$. Meanwhile, it is expected that $\langle qq \rangle$ scales as  $\propto\mu^2$ by the weak coupling analysis~\cite{Schafer:1999fe, Hanada:2011ju, Kanazawa:2013crb}. } \label{table:phase}
\end{center}
\end{table}
%%%%%%%%%%%%%
In this section, we first provide the definition of these observables and give a brief review of related recent works to obtain the phase diagram. 
Later, we will also investigate the chiral condensate, $\langle \bar{q}q \rangle$, and the topological susceptibility to see the vacuum properties for each phase.

The definition of each quantity is as follows.
Firstly, the magnitude of the Polyakov loop is given by
\beq
L= \frac{1}{N_s^3} \sum_{\vec{x}} \prod_\tau U_{4} (\vec{x}, \tau),
\eeq
which plays the role of an approximate order parameter for confinement.
As we will explain, it has been found that this quantity is often affected by severe finite-volume effects. It is also important to examine the $\bar{q}$-$q$ potential to see if confinement occurs.
Secondly, the diquark condensate,
\beq
\langle qq \rangle \equiv \frac{\kappa}{2} \langle \bar{\psi}_1 K \bar{\psi}_2^T - \psi_1 K \psi_2^T \rangle,
\eeq 
is the order parameter for superfluidity, where $K=C \gamma_5 \tau_2$. Thus, we refer to the regime with $\langle qq \rangle \ne 0$ 
as the superfluid phase.
The third quantity, namely, the quark number density,
\beq
a^3 n_q= \sum_{i} \kappa \langle \bar{\psi}_i (x) (\gamma_4 -\mathbb{I}_4) e^\mu U_{4} (x) \psi_i (x+\hat{4})  + \bar{\psi}_i (x) (\gamma_4 + \mathbb{I}_4) e^{-\mu}U_4^\dag (x-\hat{4} )\psi_i (x-\hat{4}) \rangle,\nonumber\\
\eeq
is utilized to identify a BEC-BCS crossover behavior in the superfluid phase.
This quantity is the time-like component of a conserved current and does not require a renormalization.
Also, it goes to $2N_fN_c$ in the high-density limit on the lattice.

To find the critical $\mu_c$ where the hadronic-superfluid phase transition occurs, we attempt to fit a few data for $\langle qq \rangle$ right after the diquark condensate becomes non-zero by using the following scaling law,
\beq
\langle qq \rangle =A (\mu -B)^{1/2},\label{eq:qq-ChPT-scaling}
\eeq
where $A$ and $B$ are the fitting parameters.
The exponent $1/2$ is predicted by the mean-field ChPT~\cite{Kogut:2000ek}, where the fitting parameter $B$ corresponds to $\mu_c$.  According to the ChPT analysis~\cite{Kogut:2000ek} and other chiral effective models~\cite{Ratti:2004ra,Suenaga:2022uqn,Suenaga:2023xwa}, $\mu_c$ is exactly $\mu_c/m_{\rm PS}=0.50$ at zero temeprature. So far, our previous paper~\cite{Iida:2019rah} and also other independent works~\cite{Hands:2006ve, Braguta:2016cpw, Boz:2019enj} have confirmed that this prediction is almost true even at non-zero temperature, i.e., $ \langle qq \rangle \propto (\mu -\mu_c)^{1/2}$ with $\mu_c/m_{\rm PS} \simeq 0.50$ as long as the 
temperature is sufficiently low.

In the superfluid phase, it is expected that a macroscopic number of diquark $q$-$q$ pairs form a condensate in such a way that the BEC-BCS crossover occurs as the density increases.
Here, in the BEC phase, attraction between quarks in each pair is sufficiently strong to keep the pair small, while in the BCS phase, the quark dynamics can be described by a correction due to weak attraction on free theory, namely, an instability of the Fermi surface by formation of Cooper pairs.
In QCD-like theory, thanks to asymptotic freedom, the interaction between quarks becomes weaker as they approach each other, so that the BCS phase appears in the regime of sufficiently high densities to be consistent with free theory.
To find such a regime, 
we investigate the quark number density.
Thus, if the lattice result $\langle n_q \rangle$ can be approximated by its value for free theory on the lattice~\cite{Hands:2006ve},
\beq
n_q^{\rm tree}(\mu) = \frac{4N_cN_f}{N_s^3 N_\tau} \sum_k \frac{i \sin \tilde{k}_0 [ \sum_i \cos k_i -\frac{1}{2\kappa} ]}{[\frac{1}{2\kappa} -\sum_\nu \cos \tilde{k}_\nu ]^2 +\sum_\nu \sin^2 \tilde{k}_\nu},\label{eq:nq-tree}
\eeq
where
\beq
\tilde{k}_0 = k_0 -i\mu = \frac{2\pi}{N_\tau} (n_0+1/2) -i\mu,~~~~~~~~\tilde{k}_i = k_i = \frac{2\pi}{N_s}n_i,~~~~i=1,2,3,
\eeq
then we refer to such a regime as the BCS phase.
In the continuum limit, Eq.~\eqref{eq:nq-tree} scales as $n_q^{\rm tree,cont}=N_c N_f \mu^3/(3\pi^2)$ as a function of $\mu$ in the high-density limit where the Fermi surface with radius $\mu$ is perfectly constructed.  
Furthermore, according to the weak coupling analysis~\cite{Schafer:1999fe, Hanada:2011ju, Kanazawa:2013crb}, it is expected that the diquark condensate increases as $\langle qq \rangle \propto \mu^2$ in such a high-density regime.

In distinguishing between the confined and deconfined regimes in the BCS phase as depicted in Figure~\ref{fig:schematic}, there have been some progress in recent years.
Naively, it was expected that deconfinement would occur at sufficiently high densities since the theory could be expressed by a standard perturbative QCD. Indeed, a pioneering paper by S. Hands et al.~\cite{Hands:2006ve} shows that the Polyakov loop grows with increasing density at $T=45$ MeV, which was considered to be a signal of deconfinement.  In our previous work~\cite{Iida:2019rah}, however, we found a signal of severe lattice artifacts even before the density reaches the confinement-deconfinement transition at a higher temperature, $T=80$ MeV.
Thus, in the density regime relevant for our simulations, confinement remains even in the BCS phase at $T=80$ MeV. We also found that the $\bar{q}$-$q$ potential shows a linear behavior with respect to the relative distance even at the highest density studied at $T=40$ MeV~\cite{Ishiguro:2021yxr}.
Soon after our previous paper\cite{Iida:2019rah}, S.~Hands et al.\ updated the data with finer lattice spacing and larger lattice volume simulations, which show that the renormalized Polyakov loop goes to zero around $T=100$ MeV even at a high density corresponding to $\mu \approx 700$ MeV. 
Furthermore, another group, by using staggered fermions~\cite{Begun:2022bxj}, investigated the critical temperature for deconfinement, $T_d$, in a high-density regime; $T_d\approx 100$ MeV was found even at $\mu \approx 1.8$ GeV.
In summary, the QC$_2$D system remains confined in the BCS phase at low temperature, as shown in Figure~\ref{fig:schematic}.
The theoretical analysis using 't Hooft anomaly matching for massless QC$_2$D also give a constraint $T_d \le T_{\rm SF}$ 
at fixed $\mu$, where  $T_{\rm SF}$ denotes the critical temperature for the QGP-superfluid transition~\cite{Furusawa:2020qdz}.
It is consistent with Figure~\ref{fig:schematic} based on the recent lattice results.

A combination of $\langle qq \rangle$ and $\langle n_q \rangle$ plays another role in finding the hadronic-matter phase, where $\langle qq \rangle =0$ and $\langle n_q \rangle >0$. 
This phase was found in our previous work~\cite{Iida:2019rah}  in the course of a detailed study of the vicinity of the hadronic-superfluid phase transition.
Theoretically, at exactly zero temperature, the emergence of the diquark condensate breaks quark-antiquark symmetry and causes non-zero quark number density.
At non-zero temperatures, however,  a positive $\langle n_q \rangle$ can appear even if $\langle qq \rangle =0$ due to thermal fluctuations.
Indeed, an imbalance can occur
between diquark and antidiquark excitations, since near the critical point, the scalar diquark, which is the lighest hadron in the superfluid phase, is much lighter than the scalar antidiquark.
For appearance of abundant diquarks at a given $\mu$ below $\mu_c$, the effective mass of the diquark is expected to be less than $1/(aN_\tau)=T$.
At the same $\mu$, $\langle qq \rangle$ remains zero in the ground state, but
non-zero $\langle n_q \rangle$ is thermally induced.
In fact, our previous numerical result shows that $\langle qq \rangle$ is consistent with zero at $a\mu < 0.30$, while $\langle n_q \rangle >0$ was observed at $a\mu \geq 0.26$ at $T=80$ MeV in the $j\rightarrow 0$ limit.  
We will see that the range of $\mu$ that covers the hadronic-matter phase becomes narrower at the lower temperature, $T=40$ MeV.

It might be worth while to see the chiral condensate, $\langle \bar{q} q \rangle \equiv \langle \bar{\psi}_i \psi_i \rangle$, which is well-known as an order parameter of chiral symmetry breaking~\footnote{In this definition, we consider the chiral condensate for one flavor and do not take the sum over flavor index $i$. }.  In our numerical calculations, 
however,  we utilize the massive Wilson fermions, which explicitly break the 
chirality.  Although we will discuss numerical results for the chiral 
condensate in each phase, we will not use the quantity for the definition 
of the phase.

Furthermore, we will study possible classical configurations by observing the topological-charge distribution. It is widely believed that in the high-temeprature limit, the QCD vacuum is governed by the perturbation vacuum where there is no instanton, while in the low temperature regime with $\mu=0$, non-zero instanton configurations are allowed.
A question to be addressed here is the following: Whether the vacuum is strictly perturbative or not in the high-density and low-temperature regime, where the quark dynamics has to be described by perturbative QCD.

The topological charge can be measured by incorporating the gluonic definition,
\beq
Q (t) = \frac{1}{32 \pi^2} \sum_{x} \mbox{Tr} \epsilon_{\mu \nu \rho \sigma} 
G^a_{\mu \nu} (x,t) G^a_{\rho \sigma} (x,t),
\eeq
into the gradient flow method~\cite{Luscher:2010iy}.  Here, the field strength $G_{\mu \nu}(x,t)$ is constructed by the smeared link variables at finite flow-time $t$. The field strength is calculated by using the clover-leaf operator on the lattice. 
In the gradient flow process, we utilize the Wilson-plaquette action as a kernel of the flow equation. 

The value of $Q(t)$ roughly plateaus for long $t$, but small fluctuations remain. Therefore, we introduce a reference scale $t_0$ and identify the value of $Q(t=t_0)$ as a convergent value of $Q$ for each configuration~\cite{Bruno:2014ova}.
The reference scale $t_0$ is originally introduced in Ref.~\cite{Luscher:2010iy}, it is given by
\beq
t^2 \langle E(t) \rangle |_{t=t_0} =0.3,
\eeq
where $E(t)$ denotes the energy density
\beq
E(t)= -\frac{1}{2V} \sum_x \mathrm{tr} \{ G_{\mu \nu} (x,t) G_{\mu \nu}(x,t) \}.
\eeq
A good quantity to see the distribution of the topological charge is the topological susceptibility,
\beq
\chi_Q = \langle Q^2 \rangle - \langle Q \rangle^2.
\eeq

In our previous work~\cite{Iida:2019rah}, we found that the topological susceptibility is almost constant at $T = 80$ MeV for all the values of $\mu$ studied there, which range from the hadronic to BCS phase. This is in contrast to the high-temperature regime, at $T=160$ MeV, where the topological susceptibility clearly decreases as the hadronic phase changes to the quark-gluon-plasm phase with increasing $\mu$~\cite{Iida:2019rah}.
In the previous work, we measured the topological charge for generated configurations at $aj =0.02$ $(0.00)$ in the superfluid (hadronic) phase.
In this work, we carefully investigate the topological susceptibility as a function of $\mu$ by taking the $j \rightarrow 0$ limit.

%%%%%%%%%%%%%%%%%%%%%%%%%%%%%%%%%%%%%%%%%%%%%%%%%%%%%%%%%%
%%%%%%%%%%%%%%%%%%%%%%%%%%%%%%%%%%%%%%%%%%%%%%%%%%%%%%%%%%
\subsection{Results: Phase diagram}
%%%%%%%%%%%%%%%%%%%%%%%%%%%%%%%%%%%%%%%%%%%%%%%%%%%%%%%%%%
%%%%%%%%%%%%%%%%%%%%%%%%%%%%%%%%%%%%%%%%%%%%%%%%%%%%%%%%%%
Now, let us show the simulation results.
As for confinement, we already studied the $\mu$-dependence of the $q$-$\bar{q}$ static potential by measuring the Wilson loop on the same lattice setup (see Figure~$1$ in Ref.~\cite{Ishiguro:2021yxr}).
The value of the string tension slightly decreases in the high-density regime, but the $q$-$\bar{q}$ potential still behaves as a linear potential. Therefore, we conclude that the system remains confined in the whole-$\mu$ regime studied in this work.

%%%%%%%%%%%%%%%%%%%%%%%%%%%%%%%%%%%%%%%%%%%%%%%%%%%%%%%%%%
\begin{figure}[htbp]
\centering
\includegraphics[width=0.43\textwidth]{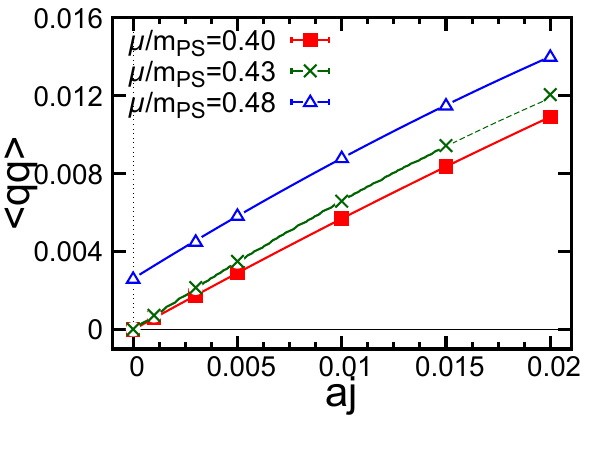}
\qquad
\includegraphics[width=.43\textwidth]{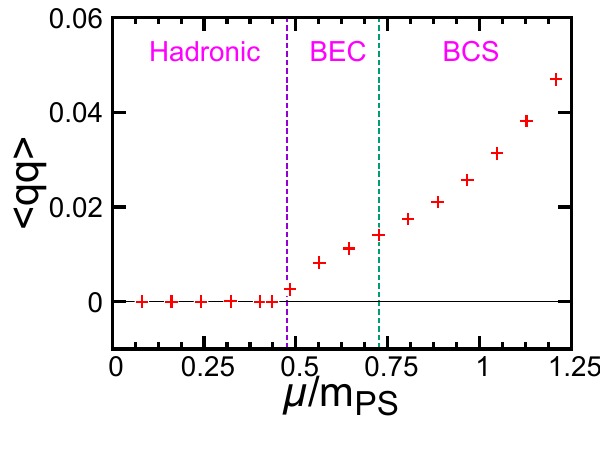}
\caption{(Left) The $aj\rightarrow 0$ extrapolation of the diqaurk condensate for $\mu/m_{\rm PS}=0.40,0.43$ and $0.48$. (Right) The $\mu$-dependence of the diquark condensate at $aj =0$.}\label{fig:j-zero-diquark-cond}
\end{figure}
%%%%%%%%%%%%%%%%%%%%%%%%%%%%%%%%%%%%%%%%%%%%%%%%%%%%%%%%%%
As for the diquark condensate, we can take a linear fitting of the $aj=0.010, 0.015$ and $0.020$ data at each $\mu$ to explore the $j \rightarrow 0$ limit, except for around $\mu/m_{\rm PS}=0.5$  where the fit quality becomes extremely bad ($\chi^2/$d.o.f. $>10$).
At $\mu/m_{\rm PS}=0.40, 0.43$ and $0.48$, therefore, we perform the same reweighting method as used in Ref.~\cite{Iida:2019rah} to obtain the data for $aj=0.001, 0.003$ and $0.005$ by using $aj=0.010$ configurations. 
In fact, we cannot directly obtain the data at $aj=0.001$ and $\mu/m_{\rm PS}=0.48$ because the solver does not converge in the calculation of the inverse of the Dirac operator.  This indicates the emergence of zero modes related to the phase transition around  $\mu/m_{\rm PS}=0.48$.

The left panel of Figure~\ref{fig:j-zero-diquark-cond} depicts the $j \rightarrow 0$ extrapolation for these values of $\mu/m_{\rm PS}$ around the critical point.
The quadratic fits of $aj$ for $5$ or $6$ data points work well~\footnote{The  datum at $aj=0.020$ and $\mu/m_{\rm PS}=0.43$ is not included to keep a reasonable fit quality.}.
 While the ChPT theory gives a prediction of how the diquark condensate scales as a function of $j$ near the critical point: $\langle qq \rangle \propto j^{1/3}$, our data look at odds with the prediction. Our previous data at $T=80$ MeV~\cite{Iida:2019rah} did not show this scaling, nor did the data obtained from Wilson fermions by S.~Hands et al.\ ~\cite{Boz:2019enj}.
 Interestingly, the study with staggered fermions  having smaller bare quark masses gave numerical evidence for the $\langle qq \rangle \propto j^{1/3}$ scaling~\cite{Braguta:2016cpw}.
The reason for such discrepancy may come from our usage of Wilson fermions with rather larger masses, which could disturb the critical behavior appreciably.

The right panel of Figure~\ref{fig:j-zero-diquark-cond} is a summary plot of the diquark condensate 
after taking the $j \rightarrow 0$ limit.
We can find that the diquark condensate starts to take non-zero values at $\mu/m_{\rm PS} = 0.48$, which is very close to the theoretical prediction by the ChPT, $\mu/m_{\rm PS}=0.50$, at  exactly zero temperature.
Here, we depict two guiding vertical lines at $\mu/m_{\mathrm{PS}}=0.47$ and $\mu/m_{\mathrm{PS}}=0.73$ $(a\mu=0.45)$  in purple and green, respectively. As we will explain below, the former $\mu$ is the critical $\mu_c$ for the hadronic-superfluid phase transition, while the latter one shows a guiding line where the BEC-BCS crossover occurs.

Now, let us closely observe the data in each phase.
%%%%%%%%%%%%%%%%%%%%%%%%%%%%%%%%%%%%%%%%%%%%%%%%%%%%%%%%%%
\begin{figure}[htbp]
\centering
\includegraphics[width=.43\textwidth]{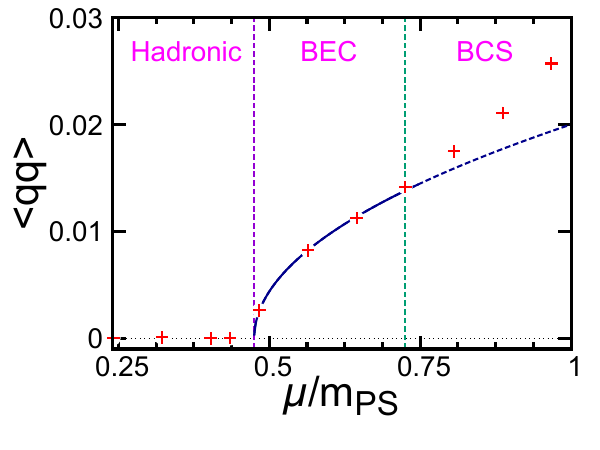}
\qquad
\includegraphics[width=.43\textwidth]{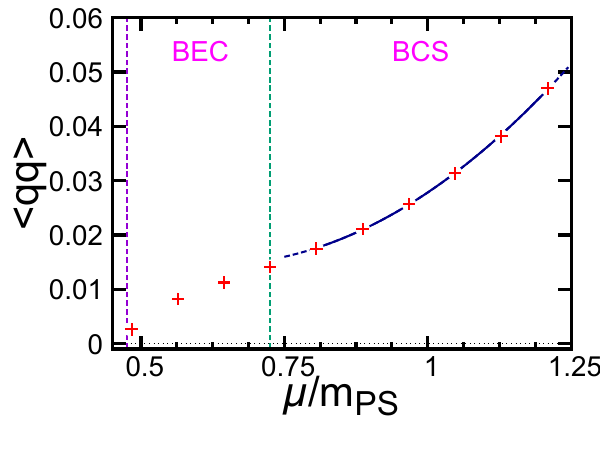}
\caption{The data for the diquark condensate in the $j\rightarrow 0$ limit as calculated for the BEC phase (left panel) and the BCS phase (right panel). The fitting functions (blue curves) are $\langle qq \rangle =A(\mu-B)^{1/2}$ in the left panel and $\langle qq \rangle = c_2 \mu^2 + c_1 \mu +c_0$ in the right panel, respectively. }\label{fig:diquark-cond-detail}
\end{figure}
%%%%%%%%%%%%%%%%%%%%%%%%%%%%%%%%%%%%%%%%%%%%%%%%%%%%%%%%%%
The left panel of Figure~\ref{fig:diquark-cond-detail} is the enlarged plot of the diquark condensate around the critical $\mu_c/m_{PS}$,  i.e., in the BEC phase.
By the fitting of $4$ data points lying in the range of $0.48 \leq \mu/m_{\rm PS} \leq 0.73$ using Eq.~\eqref{eq:qq-ChPT-scaling}, we obtain $\mu_c/m_{\rm PS}=0.47$. We refer to this value as the critical $\mu_c$ for the hadronic-superfluid phase transition.

In the regime of $\mu/m_{\rm PS} \geq 0.73$,  which we will refer to as the BCS phase, we can see the scaling  behavior of the diquark condensate with respect to $\mu$ as shown in the right panel of Figure~\ref{fig:diquark-cond-detail}.
We fit the data using a quadratic function of $\mu/m_{\mathrm{PS}}$.  
 The data clearly show the quadratic behavior, in contrast to a rather linear scaling with respect to $\mu$ suggested by our previous data in the BCS phase at T = 80 MeV. 
At the zero-temperature, the $\langle qq \rangle \sim \mu^2 $ scaling is expected by the weak coupling analysis in the high-density regime~\cite{Schafer:1999fe, Hanada:2011ju, Kanazawa:2013crb}.
Thanks to the lower temperature calculations,  we can now conclude that the thermal effects are suppressed at $T=40$ MeV and that the results approach the prediction by the zero-temperature weak-coupling analysis.

\begin{figure}[htbp]
\centering
\includegraphics[width=0.5\textwidth]{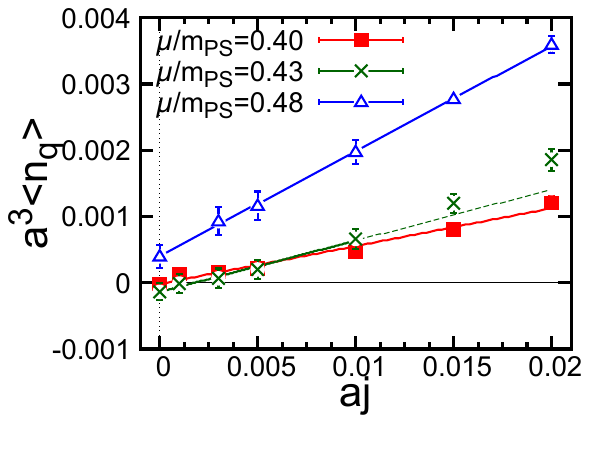}
\caption{The $aj\rightarrow 0$ extrapolation of the quark number density. In contrast to the results at $T=80$ MeV shown in the right panel of Figure~$7$ in Ref.~\cite{Iida:2019rah}, the extrapolated value (as well as the raw data at $aj=0.001$ and $0.003$) for $\mu/m_{\rm PS}=0.43$ becomes consistent with zero.}\label{fig:j-zero-nq-latt}
\end{figure}

Let us move on to the quark number density. 
The $aj \rightarrow 0$ extrapolation near the critical point is shown in Figure~\ref{fig:j-zero-nq-latt}.
%%%%%%%%%%%%%%%%%%%%%%%%%%%%%%%%%%%%%%%%%%%%%%%%%%%%%%%%%%
%%%%%%%%%%%%%%%%%%%%%%%%%%%%%%%%%%%%%%%%%%%%%%%%%%%%%%%%%%
First, we note that raw data at $aj=0.001$ both for $\mu/m_{\rm PS}=0.40$ and $0.43$ are consistent with zero. Thus,  even without the $aj\rightarrow 0$ extrapolation, we can conclude $\langle n_q \rangle =0$ for $\mu/m_{\rm PS}\leq 0.43$,  although here we dare to perform the extrapolation to do a systematic analysis. The extrapolation is done with a linear function of $aj$ for all $\mu$ regimes adopted here.

The temperature dependence of this quantity can be seen by comparison with the right panel of Figure~$7$ in Ref.~\cite{Iida:2019rah}, where even the extrapolated value for $\mu/m_{\rm PS}=0.43$ is inconsistent with zero.
Thus, as the temperature is lowered from $80$ MeV to $40$ MeV, the data for $\mu/m_{\rm PS}=0.43$ seem to change from non-zero to zero.
At $T=80$ MeV, $\mu/m_{\rm PS}=0.43$ lies in the hadronic-matter phase, i.e., $\langle qq \rangle =0$ but $\langle n_q \rangle$ is non-zero.
 
Let us quantitatively study the vanishing of the quark number density at $\mu/m_{\rm PS}=0.43$.
We recall that the hadronic-matter phase is composed of thermal diquark excitations.
If the temperature $T$ has a similar energy scale to $m_{\rm qq}$, then diquark excitations can occur appreciably.
In fact, at  $\mu/m_{\rm PS}=0.43$, we numerically obtain $am_{\rm qq}=0.0692(2)$ in the $j\rightarrow 0$ limit,  which corresponds to $m_{\rm qq} \approx 80$ MeV~\cite{Murakami-progress}.  Incidentally, this is close to the ChPT prediction given by $m_{\rm qq} = m_{\rm PS} - 2 \mu$, which results in $m_{\rm qq} = 0.14 m_{\rm PS} \approx 100$ MeV.
Thus, $T=80$ MeV can be regarded as the threshold temperature where the hadronic-matter phase can emerge due to the finite-temperature effect at $\mu/m_{\rm PS}=0.43$.   Given that $T=40$ MeV is significantly lower that the threshold temperature, it is natural that 
the hadronic-matter phase 
disappears at this temperature and $\mu$.
At a little bit higher $\mu$, namely, $\mu/m_{\rm PS}=0.48$, however,
$\langle n_q \rangle >0$ is obtained even at $T=40$ MeV, which
comes from 
the emergent superfluidity, $\langle qq \rangle \ne 0$.
Even at exactly zero temperature,  $\langle n_q \rangle >0$ remains in the superfluid phase since the baryon-antibaryon symmetry is spontaneously broken.
We expect that the $\mu$ range of the hadronic-matter phase shrinks  with decreasing temperature and eventually disappears at zero temperature.

\begin{figure}[htbp]
\centering
\includegraphics[width=.4\textwidth]{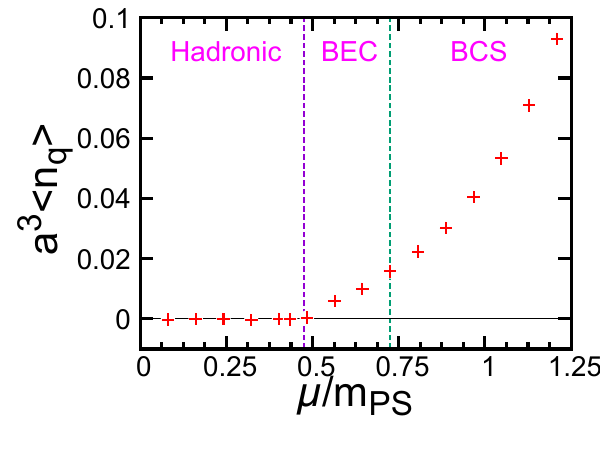}
\qquad
\includegraphics[width=.4\textwidth]{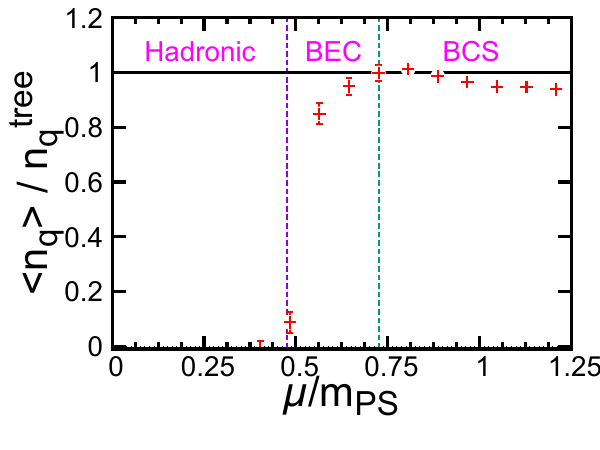}
\caption{The chemical potential dependence of the quark number density. The left panel shows $a^3 \langle n_q \rangle$,  while in the right panel, $\langle n_q \rangle$ is normalized by the free theory calculation (Eq.~\eqref{eq:nq-tree}). \label{fig:number-density}}
\end{figure}

The summary plot for the quark number density in lattice unit obtained by taking the $j\rightarrow 0$ limit is given in the left panel of Figure~\ref{fig:number-density}.
%%%%%%%%%%%%%%%%%%%%%%%%%%%%%%%%%%%%%%%%%%%%%%%%%%%%%%%%%%
%%%%%%%%%%%%%%%%%%%%%%%%%%%%%%%%%%%%%%%%%%%%%%%%%%%%%%%%%%
We can observe that the quark number density monotonically increases as $\mu$ increases in the superfluid phase.
In the right panel of Figure~\ref{fig:number-density}, the quark number density is normalized by the tree level (free quark) calculations, Eq.~\eqref{eq:nq-tree}.
The normalized quantity reaches unity at $\mu/m_{\rm PS}=0.73$ ($a\mu=0.45$), 
as is the case with $T=80$ MeV~\cite{Iida:2019rah}, and beyond this point the free particle picture looks valid. Therefore, we refer to the regime of $\mu/m_{\rm PS} \geq 0.73$ as the BCS phase, and put a guiding line at $\mu/m_{\rm PS}=0.73$ as a typical BEC-BCS crossover point.

Finally, we show the results for the chiral condensate.
Here, we utilize the Wilson fermion as a lattice fermion, so that both additive renormalization and multiplicative renormalization would be needed to discuss the chiral symmetry breaking,  whose elucidation is a hard task  to do here.   It is nevertheless interesting to  see a tendency of  how the chiral condensate depends on $\mu$.
%%%%%%%%%%%%%%%%%%%%%%%%%%%%%%%%%%%%%%%%%%%%%%%%%%%%%%%%%%
\begin{figure}[htbp]
\centering
\includegraphics[width=.5\textwidth]{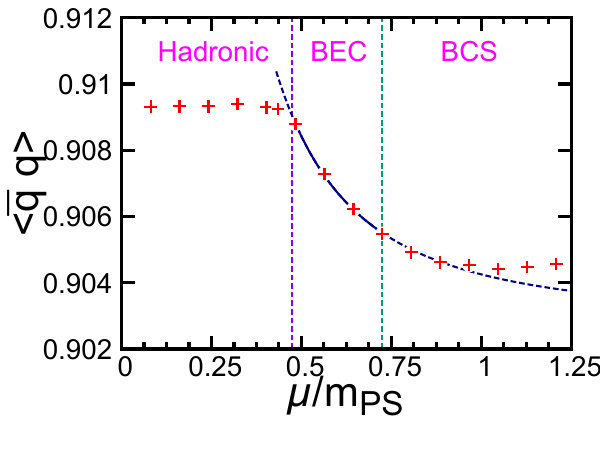}
\caption{The chemical potential dependence of the chiral condensate in the $j=0$ limit.}\label{fig:chiral-cond}
\end{figure}
%%%%%%%%%%%%%%%%%%%%%%%%%%%%%%%%%%%%%%%%%%%%%%%%%%%%%%%%%%
Figure~\ref{fig:chiral-cond} presents the result after taking $aj \rightarrow 0$ extrapolation for each $\mu$. The $aj$-dependece of the chiral condensate is very gentle and  hence the linear extrapolations work very well for all $\mu$ regimes.
We can see that the value of the chiral condensate decreases with $\mu$ in the BEC phase and reaches a constant behavior in the BCS phase. 
Thus, the chiral symmetry breaking becomes milder in the superfluid phase than in the hadronic phase.
Moreover, our data in the BEC phase is consistent with a scaling law predicted by ChPT, which is given by $\langle \bar{q}q \rangle \propto (\mu_c/\mu)^2$. In the Figure~\ref{fig:chiral-cond}, we also plot the best-fit function, $f(\mu) = c_1 (\mu_c/\mu)^2+c_0$.

The study using staggered fermions with smaller quark mass also shows a similar tendency of the $\mu$ dependence of the chiral condensate~\cite{Braguta:2016cpw,Astrakhantsev:2020tdl}. 
Furthermore, according to the analysis by 't Hooft anomaly matching, it is possible to restore the chiral symmetry in the superfluid phase in the massless limit~\cite{Furusawa:2020qdz}.

%%%%%%%%%%%%%%%%%%%%%%%%%%%%%%%%%%%%%%%%%%%%%%%%%%%%%%
%%%%%%%%%%%%%%%%%%%%%%%%%%%%%%%%%%%%%%%%%%%%%%%%%%%%%%
\subsection{Results: Topological susceptibility}
\begin{figure}[htbp]
\centering
\includegraphics[width=.4\textwidth]{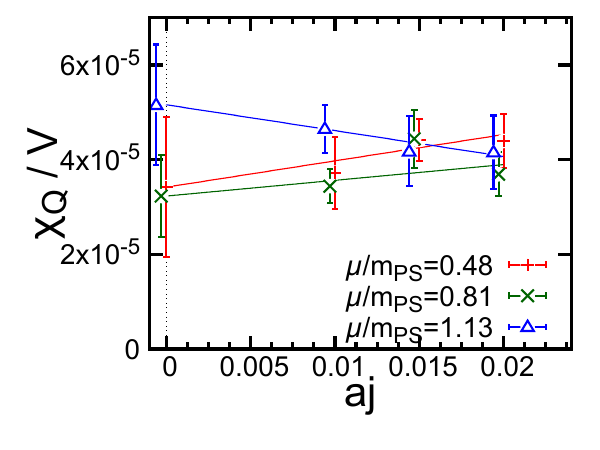}
\qquad
\includegraphics[width=.4\textwidth]{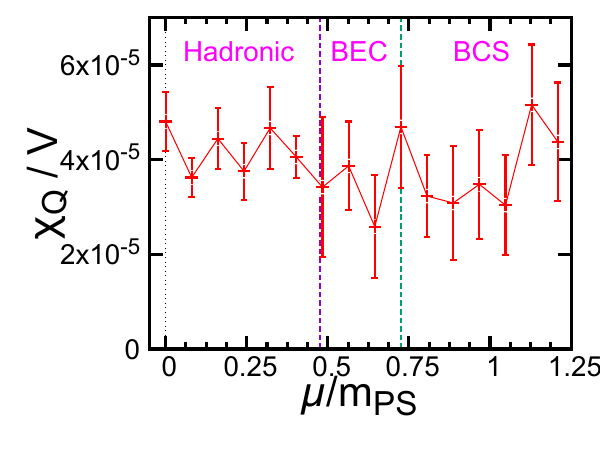}
\caption{(Left) The $aj\rightarrow 0$ extrapolation of the topological susceptibility at $\mu/m_{\rm PS}=0.48, 0.81$ and $1.13$ ($a\mu=0.30,0.50$ and $0.70$). (Right) The chemical potential dependence of the topological susceptibility in the $j=0$ limit.}\label{fig:topology}
\end{figure}

%%%%%%%%%%%%%%%%%%%%%%%%%%%%%%%%%%%%%%%%%%%%%%%%%%%%%%
%%%%%%%%%%%%%%%%%%%%%%%%%%%%%%%%%%%%%%%%%%%%%%%%%%%%%%
Finally, we present that the $\mu$-dependence of the topological susceptibility to see a nonperturbative property of generated configurations. In our previous work on $T=80$ MeV, we investigated the $\mu$-dependence of this quantity without the $aj\rightarrow 0$ extrapolation, while here we show more careful analyses.
The left panel of Figure~\ref{fig:topology} depicts the $aj$-dependence of the topological susceptibility at $\mu/m_{\mathrm{PS}}=0.48$ (near critical, bar-red symbol), $\mu/m_{\mathrm{PS}}=0.81$ (cross-green symbol), and $\mu/m_{\mathrm{PS}}=1.13$ (triangle-blue symbol).
%%%%%%%%%%%%%%%%%%%%%%%%%%%%%%%%%%%%%%%%%%%%%%%%%%%%%%%%%%
%%%%%%%%%%%%%%%%%%%%%%%%%%%%%%%%%%%%%%%%%%%%%%%%%%%%%%%%%%
No clear $aj$-dependence can be seen even at the highest density studied here. Furthermore, the extrapolated values at $aj=0$ seem non-zero and also independent of $\mu$.
The right panel of Figure~\ref{fig:topology} is a summary plot of the extrapolated values of $\chi_Q/V$ in the $j\rightarrow 0 $ limit for all $\mu$ regimes.
Just like our previous work, the results indicate that the topological susceptibility is almost constant against $\mu$; there is no decreasing behavior even in the superfluid phase. Thus, the configurations with large topological charges are generated as in the ordinary low-temperature QCD vacuum, namely, the hadronic phase.

This result is qualitatively different from what earlier investigations 
found for the $N_f=4$ theory on $12^3 \times 24$ lattice 
\cite{Hands:2011hd}, the $N_f=8$ theory on $14^3 \times 6$ lattice 
\cite{Alles:2006ea} and the recent study for the $N_f=2$ theory using staggered fermions at $T\approx 140$ MeV~\cite{Astrakhantsev:2020tdl}.
In such investigations, the topological susceptibility was found to decrease with $\mu$ in the high-density regime. 
One possible reason for this discrepancy comes from the confinement effect. In fact, in all these works, the Polyakov loop and the topological susceptibility increases and decreases simultaneously with $\mu$, while in our simulations the Polyakov loop remains close to zero even at the highest $\mu$.
In addition, in our previous work~\cite{Iida:2019rah}, we also investigated the topological susceptibility at $T\approx 160$ MeV; the data show that the topological susceptibility decreases with $\mu$ in the high-density regime where the Polyakov loop indicates the deconfinement property.
Incidentally, according to the study of finite-temperature QCD at $\mu=0$, it is well-known that the topological susceptibility is rather constant versus temperature below the deconfinement temperature and undergoes a sharp drop across the phase transition ~\cite{Alles:2000cg}.
Likewise, the present emergence of non-zero topological susceptibility even in the high-density regime might be caused by the confinement property.

On the other hand, a recent analytical study based on the linear sigma model suggests another possible explanation of the constant behavior of the topological susceptibility.
According to Ref.~\cite{Kawaguchi:2023olk}, it may come from the relatively large quark masses; indeed, the $\mu$-dependence of the topological susceptibility gets milder if the ratio of the pion (isotriplet pseudoscalar meson) and eta meson (isosinglet pseudoscalar meson) masses at $\mu=0$ is close to unity.
In our simulation, the degeneracy between the pion and eta masses at $\mu=0$ occurs~\cite{Murakami:2022lmq}.
In fact, the simulation showing the decreasing behavior was performed by using staggered fermions with relatively small quark masses~\cite{Astrakhantsev:2020tdl}.
At present, we cannot find which is more relevant for the constant behavior of the topological susceptibility, confinement or large quark masses.
It will be clarified if we perform lighter quark mass simulations.

As a summary of the phase structure of this work, we put a remark on the present data, which suggest an interesting schematic picture in the high-density regime. That is, the local quarkyonic quantities such as $\langle qq \rangle$ and $\langle n_q \rangle$ can be described by the weak-coupling theory at $T=40$ MeV, while macroscopically there remain confinement properties and also non-zero instanton configurations in the low-temperature BCS phase of QC$_2$D.

%%%%%%%%%%%%%%%%%%%%%%%%%%%%%%%%%%%%%%%%%%%%%%%%%%%%%%
%%%%%%%%%%%%%%%%%%%%%%%%%%%%%%%%%%%%%%%%%%%%%%%%%%%%%%
\section{Equation of state and sound velocity}\label{sec:eos}
%%%%%%%%%%%%%%%%%%%%%%%%%%%%%%%%%%%%%%%%%%%%%%%%%%%%%%
Now let us examine the EoS at $T=40$ MeV. In our previous letter~\cite{Iida:2022hyy} and proceedings~\cite{Itou:2022ebw, Itou:2023pcl}, we presented the selected data at $T=80$ MeV. Here, we show the detail analyses and discuss the temperature  dependence of the thermodynamic quantities and the sound velocity.

%%%%%%%%%%%%%%%%%%%%%%%%%%%%%%%%%%%%%%%%%%%%%%%%%%%%%%
\subsection{Observables and calculation strategy}
First of all, we will evaluate the $\mu$-dependence of the pressure ($p$) and internal energy density ($e$) at the fixed temperature $T=40$ MeV. 
As for the pressure, in the thermodynamic limit, it satisfies the Gibbs-Duhem relation, which allows us to calculate the pressure as
\beq
p (\mu) = \int_{\mu_c}^\mu n_q(\mu') d \mu' . \label{eq:def-p}
\eeq
Note that there is no renormalization for the quark number density as it is a conversed quantity.

On the lattice, two schemes with different discretization errors have been proposed in Ref.~\cite{Hands:2006ve}:
\beq
&\mathrm{Scheme~ I:} \frac{p}{p_{SB}}(\mu) &= \frac{\int_{\mu_o}^{\mu} d\mu' n_{q}^{latt.} (\mu') }{\int_{\mu_o}^{\mu} d\mu' n_{q}^{\mathrm{tree}} (\mu')},\\
&\mathrm{Scheme ~II:} \frac{p}{p_{SB}}(\mu) &= \frac{\int_{\mu_o}^{\mu} d\mu' \frac{n_{SB}^{cont.}}{n_{q}^{\mathrm{tree}}}   n^{latt.}_q(\mu')  }{\int_{\mu_o}^{\mu} d\mu' n_{SB}^{cont.} (\mu')}.\label{eq:p-scheme2}
\eeq
Here, $p_{SB}(\mu)$ denotes the pressure value of the non-interacting theory (the Stefan-Boltzmann (SB) limit), which is obtained by the numerical integration of the number density of quarks in the relativistic limit.
In the continuum theory, the pressure scales as $p_{SB}(\mu) = \int^\mu n_{SB}^{cont.}(\mu')d\mu' \approx N_fN_c \mu^4 /(12\pi^2)$ in the high $\mu$ regime.
Here, $\mu_o$ represents the onset scale, namely, the starting point at which $\langle n_q \rangle$ becomes non-zero as $\mu$ increases. If the thermal effect is strong, then $\mu_c \ne  m_{\rm PS}/2$ is possible. In this work we found $a\mu_o = 0.30$ ($\mu/ m_{\rm PS}=0.48$), which is close to $ \mu_c / m_{\rm PS}=0.47$ obtained in the setting of this work.

The trace anomaly basically consists of the beta-function for the parameters in the action and the trace part of the energy-momentum tensor.
How to estimate the beta-function or multiplicative renormalization factor at finite $\mu$ is nontrivial; for instance, some methods have been performed such as non-perturbative determination of Karsch coefficients~\cite{Karsch:1989fu, Hands:2011ye} or perturbative calculations~\cite{Astrakhantsev:2020tdl}.
In our previous paper~\cite{Iida:2022hyy}, we proposed the usage of non-perturbative beta-function for each parameter, which can be evaluated at $\mu=0$ along the line of constant physics (LCP). Then the trace anomaly is given by
\beq
e-3p &=& \frac{1}{N_s^3 N_\tau} \left( \left. a \frac{d \beta}{d a} \right|_{\mathrm{LCP}} \left\langle \frac{\partial S}{\partial \beta} \right\rangle_{sub.}  + \left. a \frac{d  \kappa}{d a} \right|_{\mathrm{LCP}} \left\langle \frac{\partial S}{\partial \kappa} \right\rangle_{sub.} + \left. a \frac{\partial j}{\partial a}\right|_{\mathrm{LCP}} \left\langle \frac{\partial S}{\partial j} \right\rangle_{sub.} \right). \nonumber \\  \label{eq:trace-anomaly}
\eeq
Note that the beta-function for the quark chemical potential is zero since it couples to the renormalization invariant quantity, namely, the quark number operator. 
The values of the beta-function in our lattice setup have been numerically obtained  as
 \beq
 a d\beta /da|_{\beta=0.80,\kappa=0.159}=-0.352, \quad a d\kappa/da |_{\beta=0.80,\kappa=0.159}=0.0282,\label{eq:beta-fn}
\eeq 
where we use the scale setting function (Eq.\ (23) in Ref.~\cite{Iida:2020emi})  and a set of $(\beta,\kappa)$ with a fixed mass ratio between pseudoscalar and vector mesons $m_{PS}/m_V$ in the $\mu=0$ simulations (Table~$1$ in Ref.~\cite{Iida:2020emi}).
The third term of the RHS in Eq.~\eqref{eq:trace-anomaly}, which is related to the diquark source term, is eliminated in our calculation.

In Eq.~\eqref{eq:trace-anomaly}, we take all observables, $\langle \mathcal{O} \rangle$, in the $j \rightarrow 0$ limit. 
More specifically, $\langle \mathcal{O} \rangle_{sub.} (\mu) $ denotes the quantity that has the vacuum contribution subtracted out. 
Ideally, we take $\langle \mathcal{O} \rangle_{sub.} (\mu) = \langle \mathcal{O} (\mu,T)  \rangle - \langle \mathcal{O} (\mu=0,T=0) \rangle $, but the simulation at absolute zero temperature is practically difficult.
Therefore, we perform a fixed $T$  subtraction to see the $\mu$-dependence, namely,  $\langle \mathcal{O} \rangle_{sub.} (\mu) = \langle \mathcal{O} (\mu, T\approx 40 \mathrm{~MeV})  \rangle - \langle \mathcal{O} (\mu=0, T \approx 40 \mathrm{~MeV}) \rangle $. We will leave the temperature dependence for later discussion.

The first term of the RHS in Eq.~\eqref{eq:trace-anomaly}, $\left\langle \partial S / \partial \beta \right\rangle_{sub.} $, is given by the measurement of the gauge action, where we evaluate the value of the Iwasaki gauge action for generated configurations. 
The second term essentially expresses the fermion contributions to the trace anomaly,  which can be evaluated by
\beq
\left\langle \frac{\partial S}{\partial \kappa} \right\rangle = \frac{1}{\kappa} \left( \mathrm{Tr}_{c, s,f}\mathbbm{1} - N_f \langle \bar{q}q \rangle \right).\label{eq:trace-anomaly-fermion}
\eeq
Thus, we also measure the chiral condensate, which has already been investigated in Figure~\ref{fig:chiral-cond}.

Combining Eqs.~\eqref{eq:def-p} and \eqref{eq:trace-anomaly}, we can obtain the $\mu$-dependeces of the pressure and energy density. Moreover, we can estimate the decay constant of the psuedoscalar meson ($F$), namely, the pion decay constant in QC$_2$D, using the predictions from ChPT.
According to the ChPT, the decay constant is related to these thermal quantities~\cite{Hands:2006ve}:
\beq
p_{\rm ChPT}&=&4N_f F^2 \mu^2 \left( 1- \frac{\mu_c^2}{\mu^2} \right)^2, \label{eq:ChPT-p} \\
e_{\rm ChPT}&=&4N_f F^2 \mu^2 \left( 1- \frac{\mu_c^2}{\mu^2} \right) \left( 1+3 \frac{\mu_c^2}{\mu^2} \right), \label{eq:ChPT-e} 
\eeq
near the hadronic-superfluid phase transition point.
As we have already shown, our lattice results can be well described by the ChPT predictions around the phase transition, so that the above estimation of the decay constant has to be reliable.

Finally, we investigate the sound velocity $c_{\mathrm s}^2/c^2 =\partial p/\partial e$. 
Using a symmetric difference to obtain $\partial p /\partial \mu$ and $\partial e /\partial \mu$, we evaluate 
\beq
c_\mathrm{s}^2 (\mu)/c^2= \frac{\Delta p (\mu)}{\Delta e (\mu)} = \frac{ p(\mu +\Delta \mu) - p(\mu -\Delta \mu)}{e(\mu +\Delta \mu) - e(\mu -\Delta \mu)}\label{eq:sound-velocity}
\eeq
at fixed temperature.
Note that in the standard definition, the sound velocity squared is given by $\partial p /\partial e |_{s=\mathrm{const.}}$ where $s$ denotes the entropy per baryon, but here we calculate $\partial p /\partial e |_{T=\mathrm{const.}}$.
It is technically hard to evaluate the former one in lattice simulations. 
At $T=0$, both quantities are equivalent to each other. 
Careful study of the temperature dependence of the latter one is an important task and thus will be addressed below.

%%%%%%%%%%%%%%%%%%%%%%%%%%%%%%%%%%%%%%%%%%%%%%%%%%%%%%%%%%
%%%%%%%%%%%%%%%%%%%%%%%%%%%%%%%%%%%%%%%%%%%%%%%%%%%%%%%%%%
\subsection{Results: Equation of state and decay constant}
\begin{figure}[h]
\centering
\includegraphics[width=.4\textwidth]{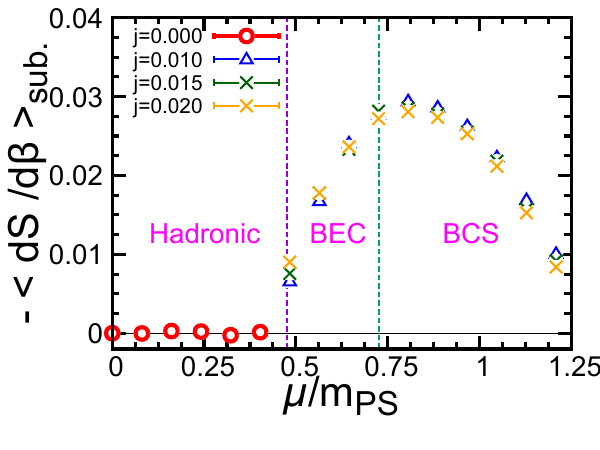}
\qquad
\includegraphics[width=.4\textwidth]{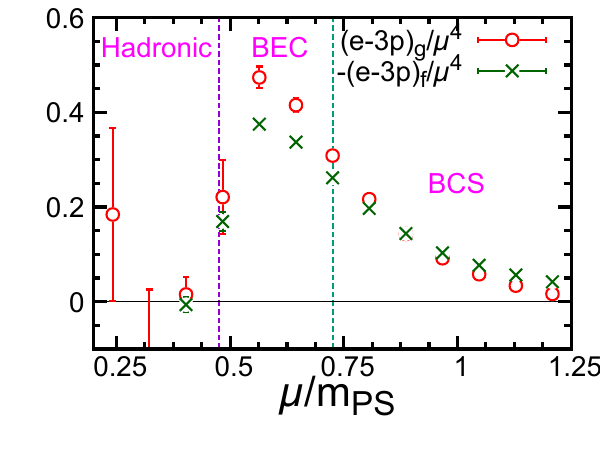}
\caption{(Left) $\langle \partial S/\partial \beta \rangle_{sub.}$ in the expression for the trace anomaly,  Eq.~\eqref{eq:trace-anomaly}.  (Right) The first (circle-red symbol) and second (cross-green symbol) terms of RHS in  Eq.~\eqref{eq:trace-anomaly}. We refer to each term  as ($e-3p$)$_g$ and  ($e-3p$)$_f$, respectively. Note that the sign of  ($e-3p$)$_f$ is flipped in the figure since it takes negative values. }\label{fig:trace-anomaly}
\end{figure}

%%%%%%%%%%%%%%%%%%%%%%%%%%%%%%%%%%%%%%%%%%%%%%%%%%%%%%%%%%
%%%%%%%%%%%%%%%%%%%%%%%%%%%%%%%%%%%%%%%%%%%%%%%%%%%%%%%%%%
The left panel of Figure~\ref{fig:trace-anomaly} shows the raw data for $\langle \partial S/\partial \beta \rangle_{sub.}$ in Eq.~\eqref{eq:trace-anomaly}. 
The different colors represent the different values of $aj$. We can see that the dependence on $aj$ is not large even in the high-density regime.
After taking the $aj \rightarrow 0 $ extrapolation by a linear function of $aj$ for each $a\mu$, the first term of RHS in Eq.~\eqref{eq:trace-anomaly},  corresponding to the gluonic contribution, has been obtained as indicated by the circle-red symbols in the right panel.
%%%%%%%%%%%%%%%%%%%%%%%%%%%%%%%%%%%%%%%%%%%%%%%%%%%%%%%%%%
%%%%%%%%%%%%%%%%%%%%%%%%%%%%%%%%%%%%%%%%%%%%%%%%%%%%%%%%%%
The cross-green symbol in the right panel of Figure~\ref{fig:trace-anomaly}, corresponding to the femionic contribution, represents the second term of RHS in Eq.~\eqref{eq:trace-anomaly}, which has been calculated by using Eq.~\eqref{eq:trace-anomaly-fermion}. This quantity is negative, so that for ease of viewing we flip the sign in the figure. 
Therefore, the red data minus the green data for each $\mu$ correspond to the trace anomaly.
We can see that in the hadronic phase, the trace anomaly is consistent with zero, which is expected from the Silver-Blaze property, while taking a positive value in the BEC phase. Finally, the magnitude of the fermion contribution (cross-green) becomes larger than the gluonic contribution (circle-red) in the high-density regime.  Indeed, the total trace anomaly takes negative values for $\mu/m_{\rm PS} \gtrsim 1.0$.

Let us consider the negativity of the trace anomaly.
One possible reason for the negativity comes from our convention where the trace anomaly at $\mu=0$ is set to zero at finite temperature ($T = 40$ MeV).
To examine this possibility, we calculate the difference between the values of  the trace anomaly at $T= 40$ MeV and $T= 80$ MeV, and then obtain
\beq
\langle e-3p ~(\mu=0, 80 {\rm ~MeV})\rangle - \langle e-3p ~(\mu=0, 40 {\rm ~MeV}) \rangle = 0.00014(220).
\eeq
This indicates that the temperature dependence is negligible and hence the negativity is robust.
Indeed, such a negative trace anomaly  in medium has been also predicted in the context of some effective  models for dense QC$_2$D and 3-color QCD with isospin chemical potential~\cite{Lu:2019diy, Kawaguchi:2024iaw}~\footnote{The trace anomaly at $\mu=T=0$ itself has been predicted to be negative owing to the structure of the QCD vacuum~\cite{Fukuda:1980py}.}.

\begin{figure}[h]
\centering
\includegraphics[width=.4\textwidth]{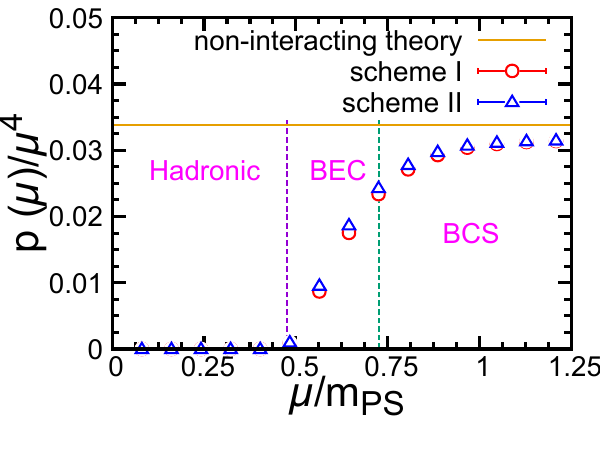}
\qquad
\includegraphics[width=.4\textwidth]{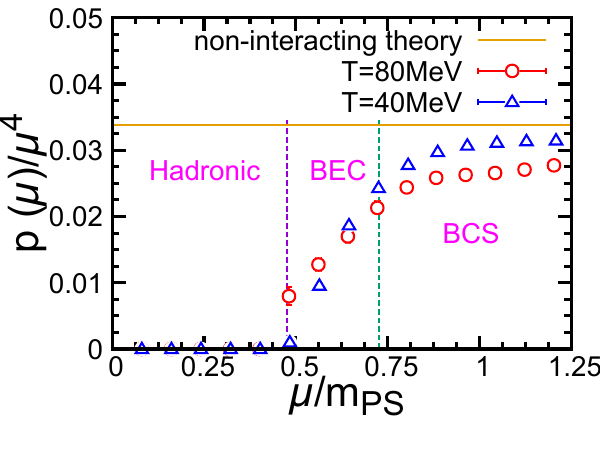}
\caption{(Left) The pressure calculated by scheme~I and scheme~II in Eq.~\eqref{eq:def-p} as a function of $\mu$. Here, the $aj \rightarrow$ extrapolation has been taken. (Right) Comparison of the pressure between $T=40$ MeV and $T=80$ MeV. Here, we plot the data in scheme~II at both temperatures. The horizontal line (orange) is the value for non-interacting theory in the continuum and thermodynamic limits, $p/\mu^4 = 1/(3\pi^2)$.} \label{fig:pressure}
\end{figure}

As for the pressure, we depict the results of both Scheme~I and Scheme~II in the left panel of Figure~\ref{fig:pressure}.
%%%%%%%%%%%%%%%%%%%%%%%%%%%%%%%%%%%%%%%%%%%%%%%%%%%%%%%%%%
%%%%%%%%%%%%%%%%%%%%%%%%%%%%%%%%%%%%%%%%%%%%%%%%%%%%%%%%%%
To obtain them, we perform the same $aj\rightarrow 0$ extrapolation with the ones for $\langle n_q \rangle$ shown in the left panel of Figure~\ref{fig:number-density} and then numerically integrate the quark number density with respect to $\mu$.
We can see from the figure that the scheme dependence is negligible, which indicates that a discretization error for this quantity is suppressed in our lattice setup. This is in contrast with Ref.~\cite{Hands:2006ve} where the values of $a$ and quark mass are different from ours.  Also, we use an improved action as the gauge action, which may cause a small discretization effect in our results.
In the high-density limit, the pressure has to scale as $p(\mu)/\mu^4 \sim  1/ (3\pi^2)$ as is the case with free theory in the continuum limit.  For comparison, we also present its value as an orange horizontal line.
Our numerical data at fixed temperature monotonically approach $1/ (3\pi^2)$; the largest values of $p/\mu^4$ are $0.03140$ (Scheme~I) and $0.03149$ (Scheme~II), which amount to $93 \%$ of the free-theory value at $T=40$ MeV. 

The right panel of Figure~\ref{fig:pressure} depicts the temperature (and volume) dependence of the pressure. Here, we  have plotted the scheme~II values in both temperatures.
In the lower temperature, the rising curve near the hadron-superfluid phase transition is steeper and hence the value in the high-density region is closer to the value of free theory in the continuum limit.
This comparison helps us to imagine that with decreasing temperature or increasing volume, the calculated pressure in the high-density region gradually would approach its behavior at exactly zero temperature and the thermodynamic limit.

\begin{figure}[htbp]
\centering
\includegraphics[width=.4\textwidth]{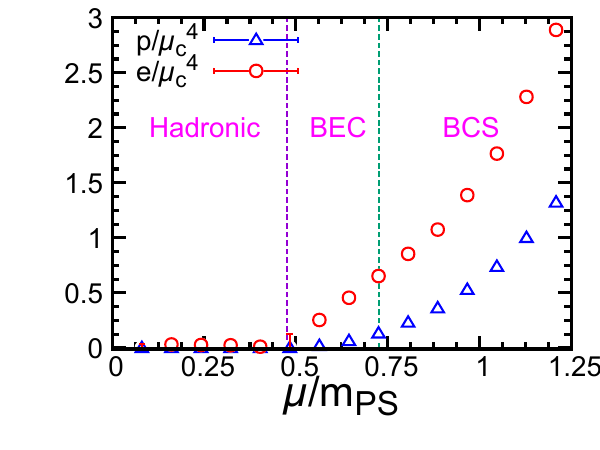}
\qquad
\includegraphics[width=.4\textwidth]{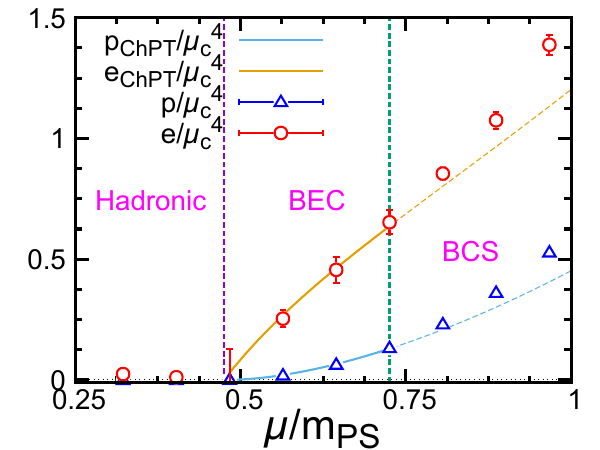}
\caption{(Left) The internal energy and pressure as a function of $\mu$. (Right) The enlarged plot of $p/\mu_c^4$ and $e/\mu_c^4$ around the BEC phase. The cyan and orange curves represent the fitting functions for $p/\mu_c^4$ and $e/\mu_c^4$, respectively, whose forms are given by 
the ChPT theory as shown in Eqs.~\eqref{eq:ChPT-p} and~\eqref{eq:ChPT-e}. }\label{fig:e-and-p}
\end{figure}

Combining the results for the trace anomaly and pressure, we obtain the energy density and pressure as a function of $\mu/m_{\rm PS}$  as shown in the left panel of Figure~\ref{fig:e-and-p}.
Here, we normalize $e$ and $p$ by using $\mu_c^4$ in such a way as to be dimensionless.
Note that both $e$ and $p$ are almost zero in the hadronic phase as they should; the former zero-consistentcy comes from the cancellation of the gluonic and fermionic terms in the trace anomaly, while the latter one simply reflects $\langle n_q \rangle =0 $.  This ensures that our calculation of the values of the beta-function~\eqref{eq:beta-fn} works well.
%%%%%%%%%%%%%%%%%%%%%%%%%%%%%%%%%%%%%%%%%%%%%%%%%%%%%%%%%%
%%%%%%%%%%%%%%%%%%%%%%%%%%%%%%%%%%%%%%%%%%%%%%%%%%%%%%%%%%

The right panel presents an enlarged plot of $p/\mu_c^4$ and $e/\mu_c^4$ around the BEC phase.
We also plot the fit functions~\eqref{eq:ChPT-p} and~\eqref{eq:ChPT-e} predicted by ChPT, where we fit the four data points $0.48 \leq \mu/m_{\rm PS} \leq 0.73$ with one fitting parameter, $F$.
The best fit values are $F= 51.1(5)$ MeV and $F=56.7(7)$ MeV obtained from the fits of $p/\mu_c^4$ and $e/\mu_c^4$, respectively.
These values are similar to the dense QC$_2$D lattice calculation result, $F=60.8(1.6)$ MeV, obtained in Ref.~\cite{Astrakhantsev:2020tdl}
from the fit of the quark number density and the mixing angle between the diquark and chiral condensates around the phase transition point predicted by the ChPT analysis.

%%%%%%%%%%%%%%%%%%%%%%%%%%%%%%%%%%%%%%%%%%%%%%%%%%%%%%
%%%%%%%%%%%%%%%%%%%%%%%%%%%%%%%%%%%%%%%%%%%%%%%%%%%%%%
\subsection{Results: Sound velocity}
%%%%%%%%%%%%%%%%%%%%%%%%%%%%%%%%%%%%%%%%%%%%%%%%%%%%%%
%%%%%%%%%%%%%%%%%%%%%%%%%%%%%%%%%%%%%%%%%%%%%%%%%%%%%%
Finally, we investigate the sound velocity squared given by Eq.~\eqref{eq:sound-velocity}.
%%%%%%%%%%%%%%%%%%%%%%%%%%%%%%%%%%%%%%%%%%%%%%%%%%%%%%%%%%
\begin{figure}[htbp]
\centering
\includegraphics[width=.6\textwidth]{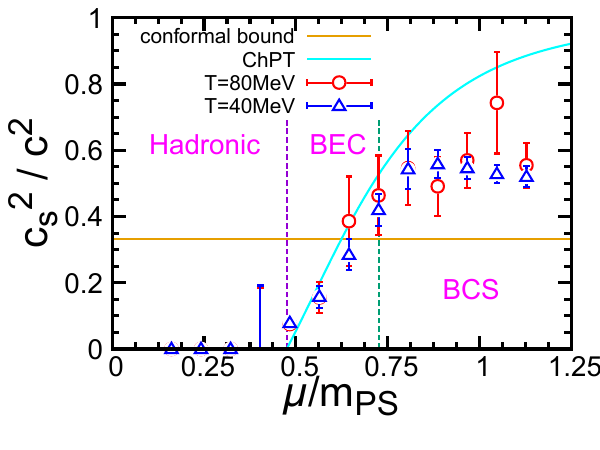}
\caption{The squared sound velocity at $T=40$ MeV and $T=80$ MeV. The cyan curve is the prediction of ChPT given by Eq.~\eqref{eq:c_s-ChPT}. The horizontal line (orange) depicts the conformal bound, $c_{\rm s}^2/c^2 = 1/3$. }\label{fig:sound-velocity}
\end{figure}
%%%%%%%%%%%%%%%%%%%%%%%%%%%%%%%%%%%%%%%%%%%%%%%%%%%%%%%%%%
In Figure~\ref{fig:sound-velocity}, the circle-red and triangle-blue data are our numerical results at $T=80$ MeV and $T=40$ MeV, respectively.
They are consistent with each other, which indicates that the temperature dependence is negligible.
The cyan curve represents the ChPT prediction~\cite{Son_2001, Hands:2006ve},  
\beq
c_{\rm s}^2/c^2=(1-\mu_c^4/\mu^4)/(1+3\mu_c^4/\mu^4),\label{eq:c_s-ChPT}
\eeq
which has no free parameter once we fix the value of $\mu_c/m_{\rm PS}$ as $0.47$.
The ChPT analysis is valid around the phase transition point, and indeed we can see that our lattice data are consistent with the prediction as expected. In the high-density regime, on the other hand, the curve given by the ChPT goes to unity,  that is, the sound velocity approaches the speed of light. Therefore, it is widely believed that the ChPT would fail at some point in the high-density regime.
Furthermore, in the high-density limit, it is believed that the EoS matches with the relativistic free theory, so that $c_{\rm s}^2/c^2$ should go to $1/3$, corresponding to $e=3p$, as shown as an orange horizontal line.
This line is called the conformal bound (or holography bound)~\cite{Cherman:2009tw}.

Our numerical results are consistent with the ChPT prediction until the sound velocity exceeds the conformal bound.
Such an excess over the conformal bound is a salient feature unknown in any lattice calculations for QCD-like theories before our previous result at $T=80$ MeV~\cite{Iida:2022hyy}~\footnote{For example, in finite temperature QCD at $\mu=0$, the sound velocity squared monotonically increases and approaches $1/3$ as the temperature increases in $T>T_c$~\cite{Borsanyi:2013bia,HotQCD:2014kol}. }.
Our new results at $T=40$ MeV obtained in this work confirm the excess over the conformal bound with a smaller statistical error. Now, the excess shows a more than $7$-$\sigma$ deviation from the conformal bound.
Furthermore, we found that the thermal effects are negligibly small,
which suggests that the difference between the definitions of $\partial p/\partial e |_{s=const.}$ and $\partial p/\partial e |_{T=const.}$ as discussed below Eq.~\eqref{eq:sound-velocity} is also negligibly small.
Then, we can safely conclude that the excess over the conformal bound in dense QC$_2$D occurs at sufficiently low temperature.

Note that the pressure itself does not exceed the free-theory limit as shown in Figure~\ref{fig:pressure}. On the other hand, the pressure growth against the energy growth, corresponding to the sound velocity, is higher than the one for the free-theory,  which supports a stiff picture for QCD(-like) matter in the superfluid regime.

%%%%%%%%%%%%%%%%%%%%%%%%%%%%%%%%%%%%%%%%%%%%%%%%%%%%%%
\subsection{Discussions: Breaking of the conformal bound and neutron star EoS}\label{sec:conformal-bound}
In this subsection, we summarize recent discussions of the conformal bound, since it has attracted attention in relation to possible evidence that could change the conventional picture of neutron stars involving first-order phase transitions.
The conformal bound (or holography bound) is a conjecture that $c_{\rm s}^2/c^2 \le 1/3$ is satisfied for a broad class of four-dimensional theories originally proposed in 2009 by Ref.~\cite{Cherman:2009tw}.
The paper itself studies the finite temperature case in the context of holography, while all lattice results for QCD and QCD-like theory at $\mu=0$ agree with the conjecture (see for instance Refs.~\cite{Borsanyi:2013bia, HotQCD:2014kol, Giudice:2017dor, Hirakida:2018uoy}). 
On the other hand, some counterexamples consisting of strongly coupled theories at finite density were also reported in the context of holography~\cite{Hoyos:2016cob, Ecker:2017fyh, Hoyos:2021uff}.

Indeed, our result in the previous work~\cite{Iida:2022hyy} was the first counterexample from the first-principles calculation. After our previous work, two independent lattice calculations for three-color QCD with isospin chemical potential ($\mu_I = \mu_u = - \mu_d $) also indicate the excess over the conformal bound in Refs.~\cite{Brandt:2022hwy, Abbott:2023coj}. 
It is well-known that theoretical structure and low-energy phenomena of three-color QCD at finite $\mu_I$ are very similar to the case of QC$_2$D at real quark chemical potential~\footnote{In both models, a color singlet condensate emerges in a high-density regime.
In fact, $3$-color QCD at non-zero $\mu_I$ is also free from the sign problem, and at $\mu_I = m_{\pi}/2$ and at sufficiently low temperature, the pion condensation emerges.}, so that these results also support our calculation results.

On the other hand, the emergence of large $c_{\rm s}$ has been pointed out by several early works based on a phenomenological quark-hadron crossover picture of neutron star matter~\cite{Masuda:2012ed,Baym:2017whm}.
It was suggested that the zero-temperature sound velocity squared, $c_{\rm s}^2=\partial p/\partial e$, peaks  as a function of $\mu$ in such as way as to be consistent with various observational constraints.
More recently, several studies based on effective models of dense QCD and dense QC$_2$D have also shown the peak structure beyond the conformal bound~\cite{McLerran:2018hbz,Fujimoto:2020tjc, Kojo:2021ugu, Kojo:2021hqh, Braun:2022olp, Braun:2022jme,Fujimoto:2022ohj,Minamikawa:2023eky,Papakonstantinou:2023myk,Carlomagno:2024xmi, Kawaguchi:2024iaw,Chiba:2024cny}.

The final question is whether such a breaking of the conformal bound is true even for real finite-density QCD.
Unfortunately, neither dense QC$_2$D nor three-color QCD at finite $\mu_I$ exactly tells us about real finite-density QCD, but such a new phenomenon itself is nevertheless very interesting.
The first-principles calculation for real finite-density QCD in a large box is still an extremely hard task.
On the other hand, it has been reported that some analyses using recent neutron star observational data 
also suggest an excess over the conformal bound~\cite{Brandes:2022nxa,Brandes:2023hma,Fujimoto:2024cyv}.
The analysis indicates evidence against a first-order phase transition, of which the occurrence has been believed for a long time~\cite{Brandes:2023hma}.

%%%%%%%%%%%%%%%%%%%%%%%%%%%%%%%%%%%%%%%%%%%%%%%%%%%%%%
%%%%%%%%%%%%%%%%%%%%%%%%%%%%%%%%%%%%%%%%%%%%%%%%%%%%%%
\section{Summary}\label{sec:summary}
%%%%%%%%%%%%%%%%%%%%%%%%%%%%%%%%%%%%%%%%%%%%%%%%%%%%%%
%%%%%%%%%%%%%%%%%%%%%%%%%%%%%%%%%%%%%%%%%%%%%%%%%%%%%%
In this paper, we investigate the phase structure and the equation of state for dense QC$_2$D at low temperatures ($T=40$ MeV, $32^4$lattice). It was previously revealed that various phases emerge below $T_c$ in the finite density region,  but the finite volume effects are sometimes large.  It is therefore important to investigate to see the temperature dependence toward zero temperature.
As a result, we have obtained roughly similar results to the previous results~\cite{Iida:2019rah, Iida:2022hyy}, but a more detailed behavior has been clarified by studying the temperature dependence.

As for the phase structure, we have found that the emergence of the superfluid phase at $\mu_c/m_{\rm PS}=0.47$, which is very close to the prediction from the ChPT phase, $\mu_c/m_{\rm PS}=0.50$.
Furthermore, there appear two phases, namely, the BEC and BCS phases, in the low and high-density regimes, respectively. 
We have shown that the scaling behavior of $\langle qq \rangle$ and thermodynamic quantities, $p, e,$ and $c_{\rm s}^2/c^2$ can be fitted in the form of the corresponding ChPT predictions around $\mu_c$.

Several new  findings have been obtained by investigating our new data at $T=40$ MeV.
First, for the hadronic-matter phase where the expectation value of the diquark condensate is consistent with zero while the quark number density is non-zero,  
it is found that the $\mu$ range that covers this phase becomes smaller at $T=40$ MeV  (this work) than at $T=80$ MeV (previous work). 
This is consistent with our expectation that such an emergent quark number density is caused by thermal excitations of the diquark, which is the lightest hadronic state in the superfluid phase.
Secondly, in the BCS phase at $T=40$ MeV, we have observed the $\langle qq \rangle \propto \mu^2$ scaling, which is predicted by the weak coupling analysis in the continuum limit.  This is in contrast to the almost linear scaling as previously found at $T=80$ MeV. 
Thirdly, we have confirmed that the topological susceptibility is non-zero and almost independent of $\mu$.  In contrast to our previous work  where we focused on the data at non-zero $aj$, here we have carefully taken the $aj\rightarrow 0$ limit.   We have pointed out the possibility that its non-zero behavior might be related to the confinement property of the BCS phase.
From these results obtained in this work, we have developed a schematic picture of the BCS  phase: The local fermionic quantities such as $\langle qq \rangle$ and $\langle n_q \rangle$ can be well-described by the weak-coupling theory, while macroscopically, nonperturbative effects manifest themselves even in the high-density and low-temperature regime where not only does confinement remain but also non-zero topological configurations are generated.

Finally, we have evaluated the EoS and $c_{\rm s}^2$ at $T=40$ MeV, which have then been compared with the previous results at $T=80$ MeV.
As $\mu$ increases, the pressure  at the lower temperature shows a more rapid growth around $\mu_c$ and  more closely approaches the free-theory value in the thermodynamic limit.  On the other hand, the temperature dependence of the sound velocity is invisible,  which allows us to use the present isothermal speed of sound in place of the usual isentropic one.  Also, breaking of the conformal bound  has been confirmed with  a smaller statistical error.

%%%%%%%%%%%%%%%%%%%%%%%%%%%%%%%%%%%%%%%%%%%%%%%%%%%%%%
%%%%%%%%%%%%%%%%%%%%%%%%%%%%%%%%%%%%%%%%%%%%%%%%%%%%%%
\appendix
\section{Raw data for EoS and sound velocity}\label{sec:raw-data}
%%%%%%%%%%%%%%
\begin{table}[h]
\begin{center}
\begin{tabular}{|c||c|c|c|}
\hline
 $a\mu$ &   $a^4 e(\mu)$                  & $a^4 p(\mu)$              & $c_{\rm s}^2/c^2$  \\
 \hline 
  0.05  &     $7.55(18.49) \times 10^{-5}$ & $0.0                   $  &      \\
  0.10  &     $2.37(1.44) \times 10^{-4} $ & $0.0                    $  &    $0.0   $    \\
  0.15  &     $2.01(76) \times 10^{-4}   $ & $0.0                    $  &    $0.0   $    \\
  0.20  &     $1.82(1.55) \times 10^{-4} $ & $0.0                    $  &    $0.0   $    \\
  0.25  &     $8.44(11.39) \times 10^{-5}$ & $0.0                    $  &    $0.05(14)$    \\
  0.30  &     $4.42(5.36) \times 10^{-4} $ & $8.19(5.64) \times 10^{-6} $  &    $0.08(1)$    \\
  0.35  &     $1.91(26) \times 10^{-3}   $ & $1.44(11) \times 10^{-4}   $  &    $0.16(3) $   \\
  0.40  &     $3.43(41) \times 10^{-3}   $ & $4.78(14) \times 10^{-4}   $  &    $0.28(4) $   \\
  0.45  &     $4.92(38) \times 10^{-3}   $ & $9.97(20) \times 10^{-4}   $  &    $0.42(4) $   \\
  0.50  &     $6.44(20) \times 10^{-3}   $ & $1.74(3) \times 10^{-3}    $  &    $0.54(6) $   \\
  0.55  &     $8.10(27) \times 10^{-3}   $ & $2.72(3) \times 10^{-3}    $  &    $0.56(4) $   \\
  0.60  &     $1.05(3) \times 10^{-2}    $ & $3.98(4) \times 10^{-3}    $  &    $0.55(3) $   \\
  0.65  &     $1.33(3) \times 10^{-2}    $ & $5.56(4) \times 10^{-3}    $  &    $0.53(2) $   \\
  0.70  &     $1.72(3) \times 10^{-2}    $ & $7.53(5) \times 10^{-3}    $  &    $0.52(3) $   \\
  0.75  &     $2.18(3) \times 10^{-2}    $ & $9.96(6) \times 10^{-3}    $  &      \\
 \hline
\end{tabular}
\caption{ Raw data  for the EoS and sound velocity at $T=40$ MeV. The pressure is calculated in the scheme~II.} 
\end{center}
\end{table}
%%%%%%%%%%%%%

%%%%%%%%%%%%%%
\begin{table}[h]
\begin{center}
\begin{tabular}{|c||c|c|c|}
\hline
 $a\mu$ &   $a^4 e(\mu)$                  & $a^4 p(\mu)$              & $c_{\rm s}^2/c^2$  \\
 \hline 
 0.05  &    $ -2.46(2.25) \times 10^{-4}$   &   $0.000                     $  &              \\
  0.10  &    $  3.07(3.27) \times 10^{-4}$   &   $0.000                     $  &   $0.0    $  \\
  0.15  &    $ -1.24(3.20) \times 10^{-4}$   &   $0.000                     $  &   $0.0    $  \\
  0.20  &    $ -2.48(4.77) \times 10^{-4}$   &   $0.000                     $  &   $0.0    $  \\
  0.25  &    $ -2.34(3.56) \times 10^{-5}$   &   $0.000                     $  &   $0.08(10)$ \\
  0.30  &    $  6.12(4.29) \times 10^{-4}$   &   $6.458(1.83) \times 10^{-5}$  &   $0.08(2) $ \\
  0.35  &    $  2.52(47) \times 10^{-3}  $   &   $1.913(155) \times 10^{-4} $  &   $0.16(5) $ \\
  0.40  &    $  3.01(38) \times 10^{-3}  $   &   $4.352(271) \times 10^{-4} $  &   $0.39(13)$ \\
  0.45  &    $  4.32(53) \times 10^{-3}  $   &   $8.735(404) \times 10^{-4} $  &   $0.46(12)$ \\
  0.50  &    $  5.38(52) \times 10^{-3}  $   &   $1.524(53) \times 10^{-3}  $  &   $0.55(11)$ \\
  0.55  &    $  7.06(41) \times 10^{-3}  $   &   $2.364(70) \times 10^{-3}  $  &   $0.49(9) $ \\
  0.60  &    $  9.22(62) \times 10^{-3}  $   &   $3.405(88)\times 10^{-3}   $  &   $0.57(8) $ \\
  0.65  &    $  1.12(7) \times 10^{-2}   $   &   $4.741(110) \times 10^{-3} $  &   $0.74(15)$ \\
  0.70  &    $  1.34(8) \times 10^{-2}   $   &   $6.500(116) \times 10^{-3} $  &   $0.55(7) $ \\
  0.75  &    $  1.85(10) \times 10^{-2}  $   &   $8.770(121) \times 10^{-3} $  &   $0.53(6) $ \\
 \hline
\end{tabular}
\caption{ Raw data for the EoS and sound velocity at $T=80$ MeV. The pressure is calculated in the scheme~II.} 
\end{center}
\end{table}
%%%%%%%%%%%%%

\section*{Acknowledgment}
We would like to thank Drs. K.~Fukushima, S.~Hands and T.~Hatsuda for helpful comments.
The work of K.~I. is supported by JSPS KAKENHI with Grant Numbers 18H05406 and 23K25864.
The work of E.~I. is supported by JST PRESTO Grant Number JPMJPR2113, %Sakigake
JSPS Grant-in-Aid for Transformative Research Areas (A) JP21H05190, %ExU
JST Grant Number JPMJPF2221  % SQAI
and also supported by Program for Promoting Researches on the Supercomputer ``Fugaku'' (Simulation for basic science: from fundamental laws of particles to creation of nuclei) and (Simulation for basic science: approaching the new quantum era), and Joint Institute for Computational Fundamental Science (JICFuS), Grant Number JPMXP1020230411. %Fugaku
The work of E.~I is supported also by Center for Gravitational Physics and Quantum
Information (CGPQI) at YITP.
E.~I and D.~S are supported by JSPS KAKENHI with Grant Number 23H05439. %Kiban-S 
K.~M. is supported in part by Grants-in-Aid for JSPS Fellows (Nos.\ JP22J14889, JP22KJ1870) and by JSPS KAKENHI with Grant No.\ 22H04917. 
The work of D.~S. is also supported by JSPS KAKENHI with Grant Number 23K03377.
The numerical simulation is supported by the HPCI-JHPCN System Research Project (Project ID: jh220021) and HOKUSAI in RIKEN.

%\paragraph{Note added.} This is also a good position for notes added after the paper has been written.

% Bibliography
\bibliographystyle{JHEP}
\bibliography{2color}

\end{document}